\documentclass[aps,prd,twocolumn,amsmath,showpacs,superscriptaddress,nofootinbib]{revtex4-1}
\usepackage{graphicx}
\usepackage{longtable}
\usepackage{float}
\usepackage{dcolumn}
\usepackage{bm}
\usepackage{appendix}
\usepackage{multirow}

\begin{document}

\title{Primordial Non-Gaussianities of inflationary step-like models}

\author{Camila P. Novaes}
\affiliation{Observat\'orio Nacional, Rua General Jos\'e Cristino 77, S\~ao Crist\'ov\~ao, 20921-400, 
Rio de Janeiro, RJ, Brazil}

\author{Micol Benetti}
\affiliation{Observat\'orio Nacional, Rua General Jos\'e Cristino 77, S\~ao Crist\'ov\~ao, 20921-400, 
Rio de Janeiro, RJ, Brazil}

\author{Armando Bernui}
\affiliation{Observat\'orio Nacional, Rua General Jos\'e Cristino 77, S\~ao Crist\'ov\~ao, 20921-400, 
Rio de Janeiro, RJ, Brazil}

\date{\today}

\begin{abstract}
We use Minkowski Functionals to explore the presence of non-Gaussian signatures in simulated 
cosmic microwave background (CMB) maps. 
Precisely, we analyse the non-Gaussianities produced from the angular power spectra emerging from 
a class of inflationary models with a primordial step-like potential. 
This class of models are able to perform the best-fit of the low-$\ell$ `features', revealed first in the 
CMB angular power spectrum by the WMAP experiment and then confirmed by the Planck collaboration 
maps. 
Indeed, such models generate oscillatory features in the primordial power spectrum of scalar 
perturbations, that are then imprinted in the large scales of the CMB field. 
Interestingly, we discover Gaussian deviations in the CMB maps simulated from the power spectra 
produced by these models, as compared with Gaussian $\Lambda$CDM maps. 
Moreover, we also show that the kind and level of the non-Gaussianities produced in these simulated 
CMB maps are compatible with that found in the four foreground-cleaned Planck maps. 
Our results indicate that inflationary models with a step-like potential are not only able to improve 
the best-fit respect to the $\Lambda$CDM model accounting well for the `features' observed in the 
CMB angular power spectrum, but also suggesting a possible origin for certain non-Gaussian signatures 
observed in the Planck data. 
\end{abstract}

%\pacs{98.80.Cq, 98.70.Vc, 98.80.Es}
\pacs{pacs}

\maketitle

\section{Introduction}

The Standard Cosmological model is efficiently supported by the inflationary paradigm to explain the 
flatness and homogeneity of the observed Universe. 
At the same time, the inflation model provides also an elegant mechanism to produce the primordial 
curvature perturbations that generate the seeds for the formation of structures. 

The most recent cosmic microwave background (CMB) data by the Planck 
satellite~\cite{PLA-XIII-2015} are in excellent agreement with the assumption of adiabatic primordial scalar perturbation with 
nearly scale-invariant power spectrum, described by a simple power law with spectral index $n_s$ 
very close to (albeit different from) unity \cite{PLA-XX-2015} .
It would be produced in the simplest inflationary scenario, in which a single, minimally-coupled scalar field slowly rolls down a smooth potential. 

In spite of this, models that accounts for the localised `features' in the primordial power spectrum 
provide a better fit to the data with respect to a smooth power-law spectrum \cite{Peiris:2003ff,Covi:2006ci,Hamann:2007pa,Mortonson:2009qv,
Hazra:2010ve,Meerburg:2011gd, Benetti:2011rp,Benetti:2012wu, Benetti:2013cja, Miranda:2013wxa}. 
Features in the primordial power spectrum can be generated following departures from slow roll, 
that can happen in more general inflationary models with a symmetry-breaking phase transition. 
It is the case of the inflationary models with a parametrized step in the primordial potential, 
which show oscillations in the power spectrum of curvature 
perturbations, localised around the scale that is crossing the 
horizon at the time the phase transition occurred~\cite{Adams:2001vc,Hunt:2004vt,Ashoorioon:2014}. 

The main tools for analyse (and constrain) the inflationary models are the current CMB temperature 
power spectrum and bispectrum~\cite{Ribeiro:2012ar,Adshead:2011jq,Martin:2011sn,Park:2012rh}. 
Another important approach to study viable inflation models is through the non-Gaussian 
signatures produced during the inflationary phase, that left imprints in the CMB temperature fluctuations. 
Recent constraints on primordial non-Gaussianities (NGs) found in the Planck Collaboration 
analyses~\cite{2013/planck-XXIV,2015/planck-XVII} severely constrain those of local type (hereafter 
{\it local}\/-NG), supporting the simplest inflationary model based on a single minimally-coupled scalar 
field. 
However, such analyses do not rule out inflationary models producing other types of primordial NG, 
whenever consistent with such constraints. 
In this scenario, inflationary models with a step-like feature in the inflaton potential deserve 
particular consideration~\cite{Chen10,Chen06}. 

Non-Gaussian signals, primordial or not, appear mixed in the CMB Planck maps, with each 
phenomenon contributing with their own signature. 
Various statistical methods are proposed in the literature~\citep{2013/planck-XXIV,%
2015/planck-XVII,1999/novikov,2010/bartolo,2014a/novaes,2014b/novaes,bernui,2010/komatsu}
in order to detect all the potential NG components in the CMB, to measure their intensity and their 
angular scale dependence. 
Here we use the Minkowski Functionals (MFs)~\cite{1903/minkowski} as statistical estimators to 
look for non-Gaussian features in several sets of simulated maps. 
Between others, we analyse Monte Carlo CMB maps, seeded by the CMB angular power spectrum originated from inflationary models with a step-like feature 
in the inflaton potential, and compare them with simulated CMB maps based on the $\Lambda$CDM 
concordance model, for which we use the current data of the Planck 
collaboration~\cite{2014/planck-XV}.

This paper is organised as follows: 
in Section~\ref{sec:model} we briefly recall the inflationary formalism and introduce the class of models 
with a step-like potential. 
In Section~\ref{sec:analysis method} we briefly introduce the MFs as statistical estimators of NGs. 
In Section~\ref{sec:results} we present the results of our analyses, and finally, 
in Section~\ref{sec:conclusion} we display our concluding remarks. 

\section{Inflationary perturbations}\label{sec:model}

In this section we first introduce the formalism to calculate the power spectrum and bispectrum from 
a given inflationary potential $V(\phi)$.
Then, we describe the departures from the slow-roll approximation and introduce the inflationary class 
of models with step.

The Lagrangian for the \textit{inflaton} scalar field $\phi$, minimally coupled to gravity is 
\begin{equation}
\mathcal{L}_{\phi}=\frac{1}{2}g^{\mu\nu}\partial_{\mu}\phi \,\partial_{\nu}\phi - V(\phi) \,, 
\nonumber
\end{equation}
with $V (\phi)$ its potential.
The equation of motion for a homogeneous mode of the field $\phi (t, x)$ is described by the 
Klein-Gordon equation
\begin{equation}
\label{eq:motion_phi}
\ddot {\phi } + 3H\dot {\phi } + \frac{dV}{d\phi }= 0 \,.
\end{equation}
This is a familiar equation for a free scalar field with an extra term ($3H \dot{\phi}$, with $H$ the 
Hubble factor) that comes from the use of the Friedmann-Robertson-Walker (FRW) metric in the Lagrangian.
The Klein-Gordon equation and the first Friedmann equation
\begin{equation}
\label{eq:fried_inflation}
H^2=\frac{8\pi }{3m_{Pl}^2}\left[\frac{1}{2}\dot{\phi }^2 + V(\phi )\right] \, ,
\end{equation}
where $m_{Pl}$ is the Planck mass,
are used to determine the background dynamics for both the Hubble parameter and the (unperturbed) inflaton field $\phi$.

In order to study the evolution of the primordial scalar fluctuations, we introduce the 
\textit{comoving curvature perturbation} $\mathcal{R}=H\frac{\delta\phi}{\dot{\phi}}$ and the 
two-point correlation function
\begin{equation}
\label{corr_function}
\langle \mathcal{R}_k \mathcal{R}_{k'} \rangle=(2\pi)^3 \delta(k+k') \mathcal{P}_\mathcal{R}(k) \, .
\end{equation}
with $\mathcal{P}_\mathcal{R}$ the curvature power spectrum.
If $\mathcal{R}$ is Gaussian, the 
Eq.~(\ref{corr_function}) depends only on the magnitude of $k$ if we use the isotropy assumption. 
If we also assume the homogeneity, we get the relation $\langle\mathcal{R}_k\rangle = 0$.

Using the gauge-invariant quantity $u=a\delta \phi$, where $a$ is the scale factor, 
we can define
\begin{equation}
%\label{eq:}
u \equiv -z \mathcal{R} \, , \nonumber
\end{equation}
with $z \equiv a\dot{\phi}/ H$.
The Fourier components of $u$ obey to the equation
\begin{equation}
\label{eq:general_equation_u}
u_k''+\left( k^2-\frac{z''}{z} \right) u_k=0 \, ,
\end{equation}
with a prime denoting a derivative with respect to conformal time $\eta$. 

The $z''/z$ evolution depends on both the dynamics of the Hubble parameter and the unperturbed 
inflaton field, governed by the Friedmann equation (Eq. (\ref{eq:fried_inflation})) and the Klein-Gordon 
equation for $\phi$ (Eq. (\ref{eq:motion_phi})). 
Integrating the Eq. (\ref{eq:general_equation_u}), we get the $u_k(\eta)$ for free-field initial 
conditions. 

The power spectrum of the curvature perturbations $\mathcal{P}_\mathcal{R}$ is related to $u$ 
and $z$ through
\begin{equation}
\label{eq:R_R_features}
\mathcal{P}_\mathcal{R}(k)=\frac{k^3}{2\pi^2} \left| \frac{u_k}{z} \right|^2, 
\end{equation}
evaluated when the mode crosses the horizon.
Assuming gaussianity and adiabaticity, this quantity contains all the necessary information for a 
complete statistical description of the fluctuations. 
Instead, for non-Gaussian fluctuations the higher-order correlations contain additional information. 
These require more statistics and are therefore more difficult to measure, especially at large 
angular scales where cosmic variance errors are significant. 
In case of NG, the fluctuations have a non-zero three-point correlation function
$\mathcal{B}_\mathcal{R}(k_1; k_2; k_3)$, named bispectrum. 
One of the simplest form of {\it local}\/-NG is described by the parameterization \cite{Baumann:2009ds}: 
\begin{equation}
\mathcal{R}(x) = \mathcal{R}_g(x) + \frac{3}{5} f_{\rm \,NL} \left[ \mathcal{R}_g(x)^2 - \langle \mathcal{R}_g(x)^2\rangle \right] \, ,
\label{eq:bispectro}
\end{equation}
where $\mathcal{R}_g$ is Gaussian. 
In this model for {\it local}\/-NG, the information is reduced to a single number $f_{\rm \,NL}$.
 
The level of NG generated during the inflation depends from the inflationary model considered.
In case one assumes a inflationary model with a single scalar field, canonical kinetic terms, initial 
adiabatic (Bunch-Davies) assumption, and slow-roll condition, then a tiny level of NG is predicted 
(see, e.g.,~\cite{Chen06}). 
Thus, a robust detection of $|f_{\rm \,NL}| > 1$ would rule out this class of models, and 
vice versa. 

Indeed, when one of these assumptions is violated, a large amount of NG is 
expected~\cite{Bartolo:2004if}. 
In this case, such models can generate a detectable and unique signal of NG of {\it local} type, 
where the bispectrum amplitude is maximized for the condition $k_1 \simeq k_2 \gg k_3$, called 
squeezed triangle configuration.

In this scenario, inflationary models with a step-like feature in the inflaton potential would have produced 
departures from Gaussianity during the inflationary phase, leaving detectable signatures in the CMB 
data~\cite{Chen10,Chen06}, as we will examine.

%----------------------------------------------------------------------------
\subsection{Inflationary step-like models}
In order to solve the Eqs. (\ref{eq:motion_phi})--(\ref{eq:general_equation_u})
is useful to introduce the \textit{slow roll} approximation,
which assumes a slowly varying inflaton field with $(1/2)\dot{\phi}^2 \ll V (\phi)$.  
In this way we can neglect the kinetic term in the Friedmann equation and the acceleration 
term in the Klein-Gordon equation.

In some inflationary models the slow roll assumption could be relaxed for a few time, 
assuming the system starts in a state where the slow roll conditions are fulfilled (in order to give 
it enough time to reach the inflationary attractor solution), and the system returns to the slow roll 
regime at a later time. 

The interruption of slow-roll leave possible detectable traces in the primordial power spectrum. 
Specifically, wavelengths crossing the horizon during this fast-roll phase will be affected, leading 
to a deviation from the usual power-law behaviour at these scales.

The brief violation of the slow-roll condition can be modelled, in a single-field model, by adding a local 
feature, such as a step \cite{Adams:2001vc} to an otherwise flat inflaton potential.
This step can be regarded an effective field theory description of a phase transition in more realistic 
multi-field models \cite{Lesgourgues:1999uc}, which may arise naturally in, e.g., 
supergravity- \cite{Adams:1997de} or M-theory-inspired inflation models \cite{Ashoorioon:2006wc}.

In this work we choose to analyse models with step-like features in the inflationary potential using 
the formalism by \textit{Adams et. al} \cite{Adams:2001vc} that add a step feature to a $V(\phi)=m^2\phi^2/2$ 
chaotic potential, i.e., by considering a potential of the form 
\begin{equation}
V(\phi) = \frac{1}{2}m^2\phi^2 \left[1+ c\tanh\left(\frac{\phi-b}{d}\right)\right] \, ,
\label{eq:Vstep}
\end{equation}
where $b$ is the value of the field where the step is located, $c$ is the height of the step and $d$ 
its slope. 

\begin{figure}[!h]
\includegraphics[scale=0.7]{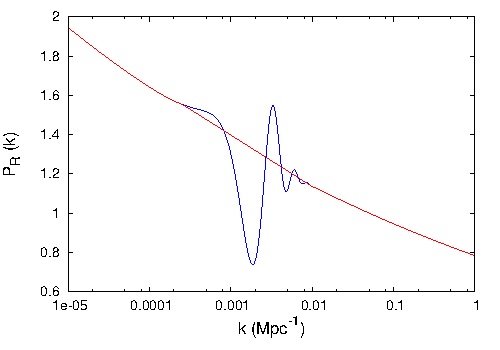}
\caption{\small Primordial power spectra for inflationary potentials with step. 
The red line shows the reference power-law spectrum, while the blue line draws the best-fit spectrum 
for the inflationary step-like model with $m = 7.5$ x $10^{-6}$, using the best-fit values 
of~\cite{Benetti:2013cja}.} 
\label{fig1}
\end{figure}

The spectrum of primordial perturbations resulting from the potential in Eq. (\ref{eq:Vstep}) can be 
calculated as outlined in the previous section, and is found to be essentially a power-law with 
superimposed oscillations, as showed in Fig. (\ref{fig1}). 
The oscillations are localized only in a limited range of wave-numbers (centred on a value that 
depends on $b$) so that asymptotically the spectrum recovers the familiar $k^{n_s-1}$ form 
typical of slow-roll inflationary models. 

The signatures of this class of models on the CMB temperature power spectrum and 
bispectrum~\cite{Benetti:2011rp, Benetti:2012wu, Benetti:2013cja, Meerburg:2011gd, Adshead:2013zfa,%
Adshead:2011jq, Adshead:2011bw, Bartolo:2013exa, Hazra2013, Sreenath2015} 
and the tensor spectrum \cite{Miranda:2014fwa, Miranda:2014wga}
were studied in detail using the current data, showing that these models improve the $\Delta \chi^2$ 
values with respect to the $\Lambda$CDM model. 
A similarly featured potential was also implemented in other context (see, e.g.,~\cite{Cai:2014vua,Jain:2008dw,Jain:2009pm}). 
Here we go beyond these results, looking at the non-Gaussian signatures of these inflationary models 
imprinted on the CMB maps. 

%------------------  FIGURE  1  ------------------------------%
%------------------  FIGURE  1  ------------------------------%

%--------------------------------------------------------------------%
\section{Analysis method}\label{sec:analysis method}

We performed a set of analyses using four Minkowski Functionals in order to search for non-Gaussian 
contributions potentially present in simulated and Planck CMB maps. 
We are primarily interested in answering the following related questions:
(i) are there non-Gaussian contributions left in simulated CMB maps generated by the angular power 
spectra of inflationary step-like models?
If the answer is yes, our second question is: 
(ii) are these non-Gaussian signatures compatible with the last released foreground-cleaned Planck CMB maps? 

In fact, if one finds non-Gaussian contributions in simulated maps produced from the CMB power 
spectra obtained from inflationary step-like models, then we can infer that these NGs are of primordial 
origin. 
For this the pertinency of the second question, to discover whether these non-Gaussian features 
are present, or not, in the foreground-cleaned Planck maps.

%--------------------------------------------------------------------o
\subsection{The Models}

We start our analysis from the results of the previous work of \textit{Benetti} \cite{Benetti:2013cja}, 
using the following values for the cosmological and step parameters: 
$100\,\Omega_b h^2=2.22$, $\Omega_{c} h^2=0.1212$, $\tau=0.089$, $100\, \theta=1.0411$, 
$n_s=0.959$, $10^9 A_s=2.20$, $b=14.66$, 
%$\log c=-2.85, \log d=-1.44$ 
$c = 1.4 \times 10^{-3}, \, d = 3.6 \times 10^{-2}$, 
and we called this model as {\em Model A}. 

We also consider a second model, called  {\em Model B}, using the same cosmological and step 
parameters values of the previous model with the exception of the step amplitude parameter 
$c$, that is four times higher than in the {\em Model A}. 

We performed a Monte Carlo Markov Chain analysis via the publicly available package 
\verb+CosmoMC+ \cite{Lewis:2002ah} and used a modified version of the \verb+CAMB+ 
(\cite{camb}) code in which we numerically solve Eqs. 
(\ref{eq:motion_phi})--(\ref{eq:general_equation_u}) using a Bulirsch-Stoer algorithm in order 
to theoretically calculate the initial perturbation spectrum Eq.~(\ref{eq:R_R_features})  
needed to compute the CMB anisotropies spectrum.
The resulting CMB power spectra for the {\em Model A} (dashed line) and {\em Model B} 
(dot-dashed line) are showed in Figs.~(\ref{fig2}) and (\ref{fig3}), together with the featureless 
$\Lambda$CDM angular power spectrum (continuous line).

%-------------------------------------------------o
\subsection{The Minkowski Functionals}

The Minkowski Functionals (MF) present many advantages as compared with other statistical 
estimators, between them one can mention their versatility to detect diverse types of NG without a 
previous knowledge of their features, like their intensity or angular dependance, and also because 
they can be efficiently applied in sky patches or masked maps. 

All the morphological properties of a $d$-dimensional space can be described using $d+1$ 
MFs~\citep{1903/minkowski}. 
In the case of a 2-dimensional CMB temperature field defined on the sphere, 
$\Delta T = \Delta T(\theta,\phi)$, with zero mean and variance $\sigma^2$, this tool provides a 
test of non-Gaussian features by assessing the properties of connected regions in the map. 
Given a sky path ${\cal P}$ of the pixelized CMB sphere ${\cal S}^2$, an {\em excursion set} of 
amplitude $\nu_t$ is defined as the set of pixels in ${\cal P}$ where the temperature field exceeds 
the threshold $\nu_t$, that is, the set of pixels with coordinates $(\theta,\phi) \in {\cal P}$ such that 
$\Delta T(\theta,\phi) / \sigma \equiv \nu > \nu_t$. 
Each excursion set, or connected region, $\Sigma$, with $\nu > \nu_t$, and its boundary, 
$\partial\Sigma$, can be defined as 
\begin{eqnarray}
\Sigma & \equiv & \{(\theta, \phi) \in \mathcal{P} ~|~ \Delta T (\theta, \phi) > \nu	\sigma_0 \}, \\
\partial\Sigma & \equiv & \{(\theta, \phi) \in \mathcal{P} ~|~ \Delta T (\theta, \phi) = \nu \sigma_0 \}.
\end{eqnarray}
In a two-dimensional case, for a region $\Sigma \subset {\cal S}^2$ with amplitude $\nu_t$ the 
partial MFs calculated just in $R_i$ are: $a_i$, the Area of the region described by 
$\Sigma$, $l_i$, the Perimeter, or contour length, $\partial\Sigma$ of this region, and $n_i$, the 
number of holes inside $\Sigma$. 
The global MFs are obtained calculating these quantities for all the connected regions with height 
$\nu > \nu_t$. 
Then, the total Area $A(\nu)$, Perimeter $L(\nu)$ and Genus $G(\nu)$ 
are~\citep{1999/novikov, 2006/naselsky_book, 2012/ducout} %, 2003/komatsu

\begin{eqnarray} \label{funcionais}
\! A(\nu) &=& \frac{1}{4 \pi} \int_{\Sigma} d\Omega = \sum a_i \, , \\
\! L(\nu) &=& \frac{1}{4 \pi} \frac{1}{4} \int_{\partial\Sigma} dl = \sum l_i \, , \\
\! G(\nu) &=& \frac{1}{4 \pi} \frac{1}{2 \pi} \int_{\partial\Sigma} \kappa dl = \sum g_i = 
N_{hot} - N_{cold} \, ,
\end{eqnarray}

\noindent
where $d \Omega$ and $dl$ are the elements of solid angle and line, respectively. 
In the Genus definition, the quantity $\kappa$ is the geodesic curvature (for details see, 
e.g.,~\cite{2012/ducout}). 
This last MF can also be calculated as the dif\/ference between the number of regions with 
$\nu > \nu_t$ (number of hot spots, $N_{hot}$) and regions with $\nu < \nu_t$ (number of cold 
spots, $N_{cold}$).

The calculations of the MFs used here were done using the algorithm developed 
by~\cite{2012/ducout} and~\cite{2012/gay}. 
This code calculates four quantities, namely $V_0 = A(\nu)$, $V_1 = L(\nu)$, $V_2 = G(\nu)$, and 
$V_3 = N_{clusters}(\nu)$ which are the three usual MFs defined above plus an additional quantity 
called \textit{number of clusters}, $N_{clusters}(\nu)$. 
The latter quantity is the number of connected regions with height $\nu$ greater (or lower) than the 
threshold $\nu_t$ if it is positive (or negative), i.e., the number of hot (or cold) spots of the map.

%----------------------------------------------------------------------------%
\subsection{Data description} \label{sec:data}

The five sets of simulated maps we investigate here were produced with 
${\rm \,N}_{\mbox{\small side}} = 512$, and in the case of the Planck CMB maps, they were degraded 
to this resolution, using the HEALPix (Hierarchical Equal Area iso-Latitude Pixelization) pixelization 
grid~\citep{2005/gorski}. 
In this section we give more details about these sets of synthetic maps as well as the Planck maps 
in analysis.

%----------------------------------------------------------------o
\subsubsection{The Monte Carlo CMB maps}

\begin{figure}[!h]
\mbox{\hspace{-0.4cm}
\includegraphics[scale=0.38]{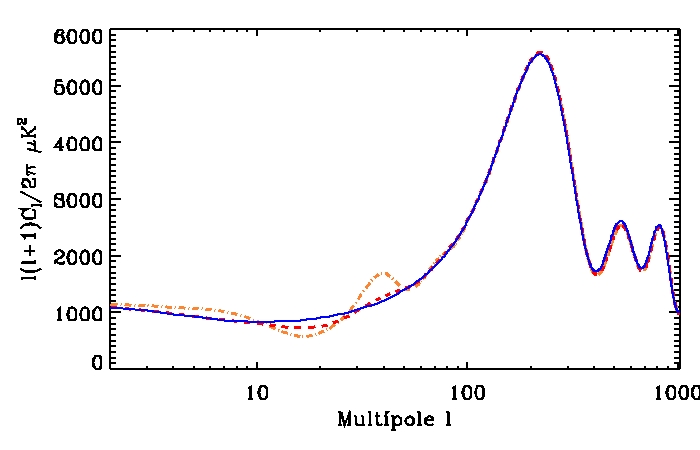}
}
\caption{\small Angular power spectra seeding the spectra of the sets of simulated CMB maps 
analysed here: 
the $\Lambda$CDM (solid blue line), the {\em Model A} (dashed red line; see text for details), and 
the {\em Model B} (dot-dashed orange line; see text for details).} 
%\label{fig:full_spectra}
\label{fig2}
\end{figure}

For our analyses we produce five sets of Monte Carlo (MC) CMB maps, random realisations considering 
a given CMB angular power spectrum. 
Each set is composed by $1000$ CMB maps. 
The first three MC sets are seeded by: 
(i) the $\Lambda$CDM angular power spectrum, according to last Planck results (solid line in 
Fig.~\ref{fig2}); 
(ii) {\em Model A}, corresponding to the best-fit of Planck angular power spectrum data using an 
inflationary model with step-like potential (dashed line in Fig.~\ref{fig2}); 
(iii) {\em Model B}, which also uses an inflationary model with step-like potential but fits less 
accurately the Planck data (dot-dashed line in Fig.~\ref{fig2}).

%------------------  FIGURE  2  ------------------------------%
%------------------  FIGURE  2  ------------------------------%

\noindent 
The last two sets of MC CMB maps are also generated using the $\Lambda$CDM power 
spectrum, but including different contributions of {\it local}\/-NG, namely, 
(iv) with $f_{\rm \,NL} = 38$, and 
(v) with $f_{\rm \,NL} = 100$, the last set.
The first one is chosen to be consistent with recent constraints found by 
Planck~\cite{2013/planck-XXIV,2015/planck-XVII}, that is, $f_{\rm \,NL} = 38 \pm 18$ (at 68\% 
confidence level) for the large angular scales. 
Moreover, our choice for $f_{\rm \,NL} = 100$ aims to test also a {\it local} non-Gaussian contribution 
with higher amplitude in order to compare the signatures they produce in the MFs.
These latter two sets of non-Gaussian CMB maps were generated from a set of coef\/ficients 
{$a_{\ell \, m}$} derived from a combination of the multipole expansion coef\/ficients 
$\{ a^{\mbox{\footnotesize G}}_{\ell \, m} \}$ and $\{ a^{\mbox{\footnotesize NG}}_{\ell \, m} \}$ 
corresponding to CMB Gaussian and non-Gaussian (of {\it local} type) maps, respectively. 
A set of each kind of these spherical harmonics coef\/ficients, 1000 linear and 1000 non-linear, 
were produced by~\cite{2009/elsner}, 
and are publicly available\footnote{http://planck.mpa-garching.mpg.de/cmb/fnl-simulations/}. 
We combine these data as follows 
\begin{equation} \label{alms_comb}
\, a_{\ell \, m} = a^{\mbox{\footnotesize G}}_{\ell \, m} 
+ f_{\rm \,NL} \, a^{\mbox{\footnotesize NG}}_{\ell \, m} \, ,
\end{equation}
which were then normalised by the Planck best fit power spectra \citep{2014/planck-XV}, rescaling 
$a^{\mbox{\footnotesize G}}_{\ell \, m}$ by the ratio of the square root of the power spectra 
and $a^{\mbox{\footnotesize NG}}_{\ell \, m}$ directly by the ratio of the power spectra. 
This linear combination permits to derive CMB synthetic maps with an arbitrary level of of {\it local}\/-NG 
defined by any real value of $f_{\rm \,NL}$.

As can be observed in Fig.~\ref{fig2}, the main dif\/ferences between the $\Lambda$CDM 
power spectrum and those from {\em Models A} and {\em B} are mainly concentrated at large scales, 
specifically for $\ell = 10 - 50$. 
For this reason all our analysis will be done focusing on this range of the angular power spectra. 
That is, in all the MC CMB maps here analysed we consider only the angular scales of this range 
of multipoles, i.e., $\ell = 10 - 50$ (see Fig.~\ref{fig3}).

%----------------------------------------------------------------o
\subsubsection{The foreground-reduced Planck CMB data}

In February 2015, the Planck Collaboration released the second set of products derived from 
the full Planck dataset~\footnote{
Based on observations obtained with Planck (http://www.esa.int/Planck), an ESA science mission 
with instruments and contributions directly funded by ESA Member States, NASA, and Canada.}. 
Between them are the four CMB foreground-cleaned maps, namely Spectral Matching Independent 
Component Analysis ($\mathtt{SMICA}$), Needlet Internal Linear Combination ($\mathtt{NILC}$), 
Spectral Estimation Via Expectation Maximization ($\mathtt{SEVEM}$), and the Bayesian 
parametric method $\mathtt{Commander}$, names originated from the separation method used 
to produce them~\citep{2015/planck-IX}. 
The performance of the foreground-cleaning algorithms has been carefully investigated by the 
Planck collaboration with various sets of simulated maps. 
These high resolution maps have ${\rm \,N}_{\mbox{\small side}} = 2048$ and effective beam 
${\rm FWHM} = 5$ arcmin.

In particular, each component separation algorithm processed dif\/ferently the multi-contaminated 
sky regions obtaining, as a consequence, dif\/ferent final foreground-cleaned regions for each 
procedure. 
These regions are defined outside the so-called Component Separation Confidence masks 
(brief\/ly termed CS-masks), which have $f_{sky} = 0.85, 0.96, 0.84, 0.82,$ for the 
$\mathtt{SMICA}$, $\mathtt{NILC}$, 
$\mathtt{SEVEM}$, and $\mathtt{Commander}$ maps, respectively. Each CMB map was released 
with its own CS-mask, outside which the corresponding CMB signal is considered statistically robust. 
Regarding the masks, it is worth mentioning that the \textit{Planck} team produced the $\mathtt{UT78}$ 
mask, which is the union of the above mentioned masks, a more restrictive one, since it has 
$f_{sky} \simeq 0.78$, and adopted as the preferred mask for analysing the temperature 
maps~\citep{2015/planck-IX}. 

\begin{figure}[!h]
\includegraphics[scale=0.40]{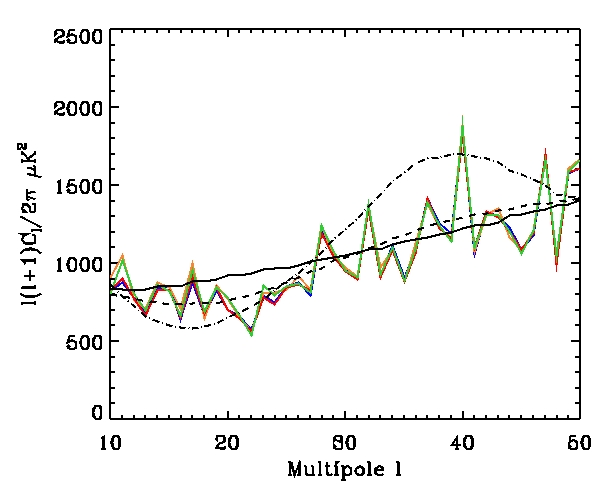}
\caption{The multipoles range of the seeded angular power spectra for the MC CMB maps, 
namely, $\Lambda$CDM (solid line), {\em Model A} (dashed line), and {\em Model B} (dot-dashed 
line). 
The blue, orange, red, and green curves correspond to the spectra calculated from the 
$\mathtt{SMICA}$, $\mathtt{SEVEM}$, $\mathtt{NILC}$ and $\mathtt{Commander}$ maps, 
respectively. 
}
\label{fig3}
\end{figure}

All the analyses here are performed upon the products of the Planck's second data release. 
Moreover, we analyse both the simulated and the four foreground-cleaned Planck CMB maps using 
two masks, namely, $\mathtt{UT78}$ and the $\mathtt{Commander}$ CS-mask (hereafter, 
$\mathtt{Commander}$-mask). 
Since the interesting region of the angular power spectrum for the current analysis corresponds to 
$\ell = 10 - 50$ (see Fig. \ref{fig3}), we performed some preliminary steps upon each Planck CMB 
map before proceeding. 
These steps are: 
(1) to degrade the maps to ${\rm \,N}_{\mbox{\small side}} = 512$, 
(2) to calculate the angular power spectrum of each one, and 
(3) to generate the CMB maps from 
the selected range of multipoles. 
These final maps are the ones we analyse in the next section.

%------------------  FIGURE  3  ------------------------------%
%------------------  FIGURE  3  ------------------------------%

%----------------------------------------------------------------o
\section{Results and Discussion \label{sec:results}}

Considering $n$ dif\/ferent thresholds $\nu = \nu_{_{1}}, \nu_{_{2}}, ...\, \nu_n$, previously defined 
dividing the range $- \nu_{max}$ to $\nu_{max}$ in $n$ equal parts, we compute the four MFs 
$\{V_k, \, k=0,1,2,3 \} \equiv (V_0,V_1,V_2,V_3)$ for the $i$-th simulated map, with $i = 1, 2,..., 1000$. 
Then, for the $i$-th map and for $k$-th MF we have the vector 
\begin{eqnarray} \label{def_v}
\,\, \mathrm{v}_k^i \equiv ( V_k(\nu_{_{1}}),V_k(\nu_{_{2}}), ... V_k(\nu_n) )|_{\mbox{\small for the 
$i$-th map}} \, .
\end{eqnarray}
In this work the values chosen for such variables are 
$\{\nu_{max},n\} = \{3.5, 26\}$ (for details, see~\citep{2012/ducout}).

%--------------------------------------- A ---------------------------------------------o
\subsection{Performance of the MFs as non-Gaussian estimators}

Firstly we investigate the performance of the four MFs by quantifying how large (in relative dif\/ference) 
is the deviation signal in the presence of {\it local}\/-NG in simulated maps, as compared with the results 
obtained in the analysis of the set of Gaussian $\Lambda$CDM CMB maps. 
For this reason we consider sets of simulated CMB maps with the same type of {\it local}\/-NG, but 
different amounts of such contribution, namely with $f_{\rm \,NL} = 38$ and $f_{\rm \,NL} = 100$, 
sets (iv) and (v), respectively. 
A similar situation happens for the MC sets (ii) and (iii), corresponding to {\em Models A} and {\em B}, 
respectively.

\begin{figure*}[!h]
\includegraphics[scale=0.33]{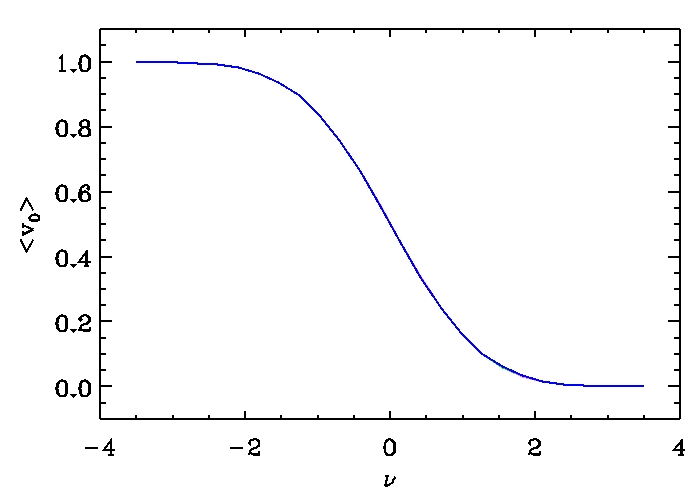}
\includegraphics[scale=0.33]{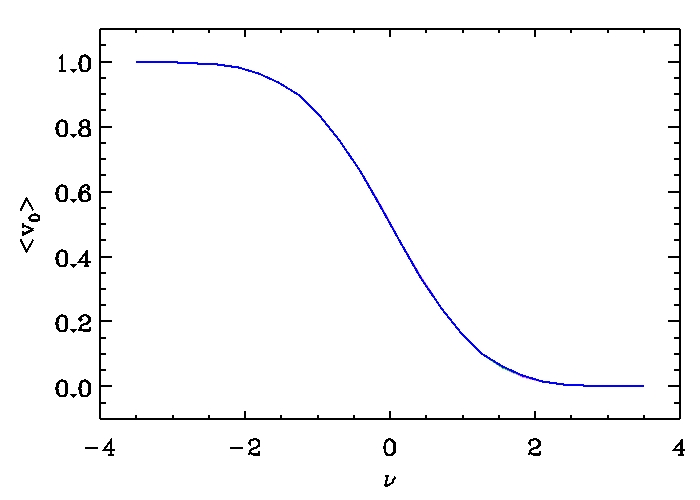} \\
\includegraphics[scale=0.33]{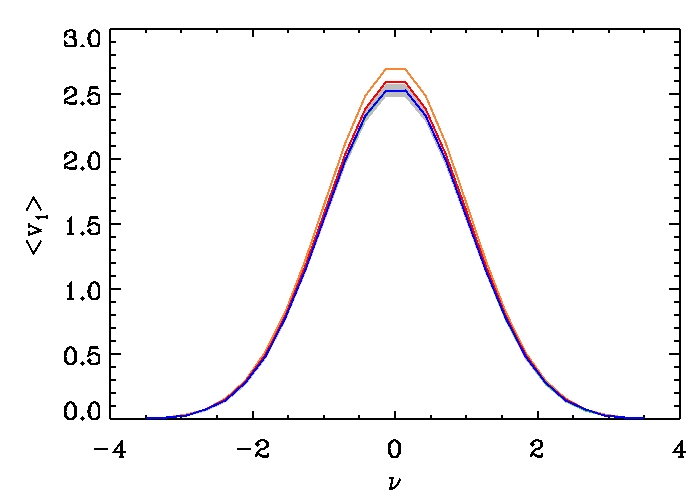}
\includegraphics[scale=0.33]{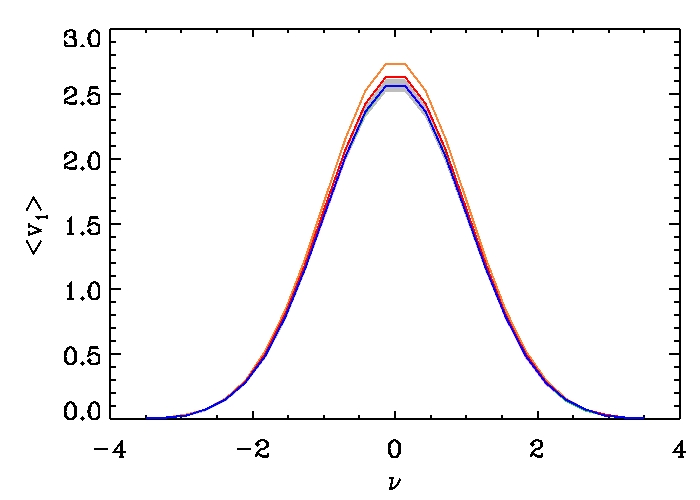} \\
\includegraphics[scale=0.33]{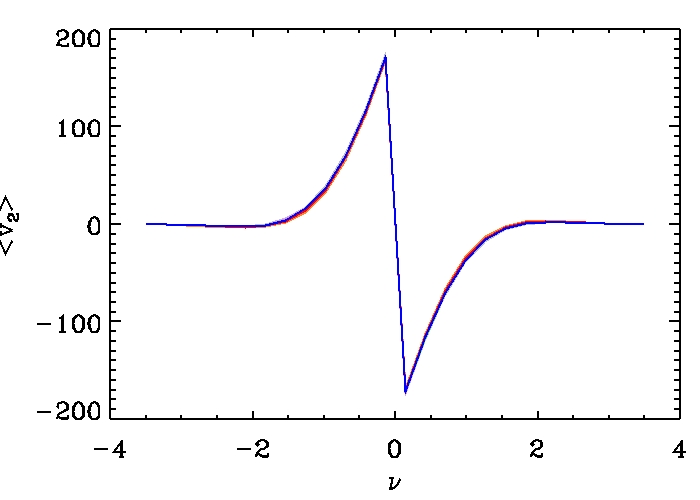}
\includegraphics[scale=0.33]{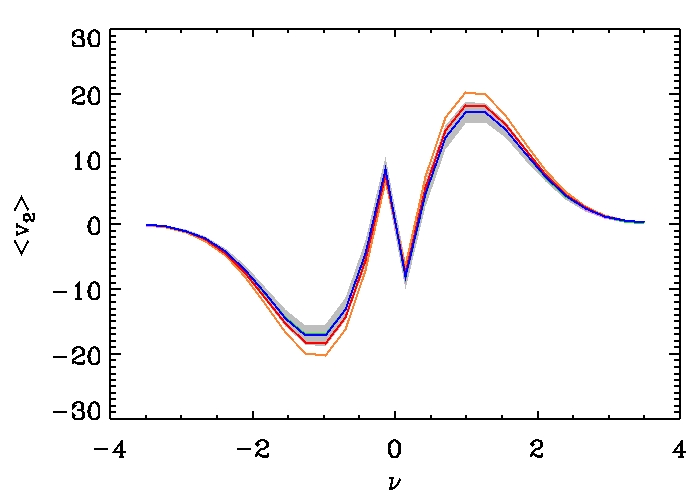} \\
\includegraphics[scale=0.33]{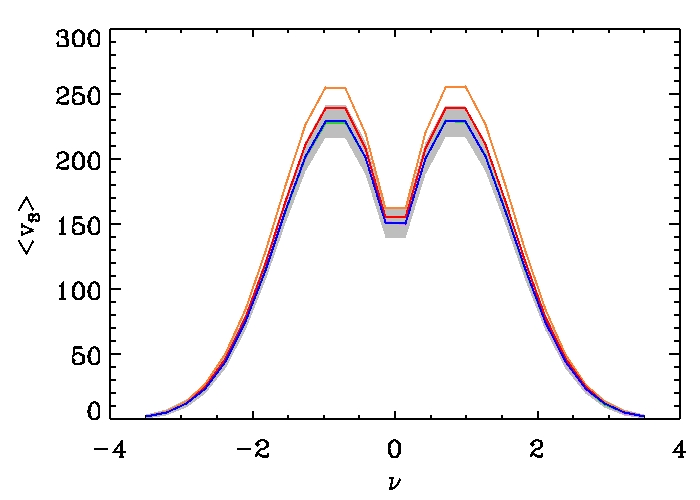}
\includegraphics[scale=0.33]{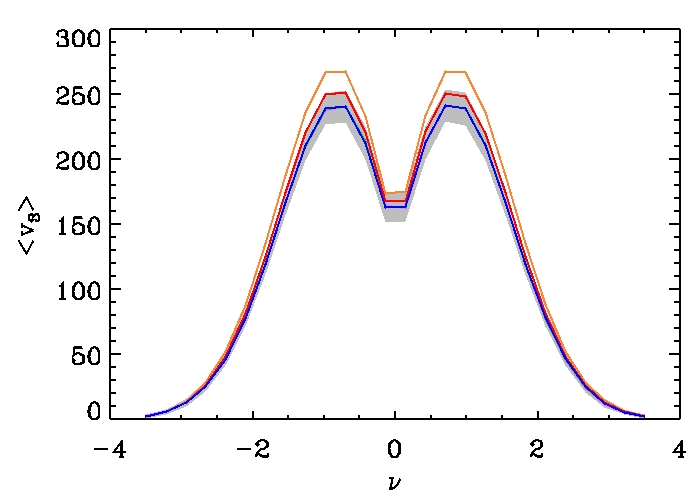}
\caption{\small Comparison between the average Area ($\mathrm{v}_0$), Perimeter ($\mathrm{v}_1$), 
Genus ($\mathrm{v}_2$) and, $N_{clusters}$ ($\mathrm{v}_3$) MF vectors over 1000 realisations of 
each dataset, using the two masks, $\mathtt{UT78}$ (left) and $\mathtt{Commander}$-mask (right). 
The blue, red, and orange curves corresponds to the average MF obtained from dataset 
seeded by the $\Lambda$CDM, {\em Model A} and {\em Model B} power spectra, respectively. 
The green and violet curves represent the average MFs calculated from MC 
CMB maps seeded by $\Lambda$CDM power spectra and contaminated by {\it local}\/-NG with 
$f_{\rm \,NL} = $ 38 and 100, respectively. The gray region represents the 1$\sigma$ level calculated 
from the MFs curves obtained from the $\Lambda$CDM seeded dataset.} \label{fig4}
\end{figure*}

\begin{figure*}[!h]
\includegraphics[scale=0.31]{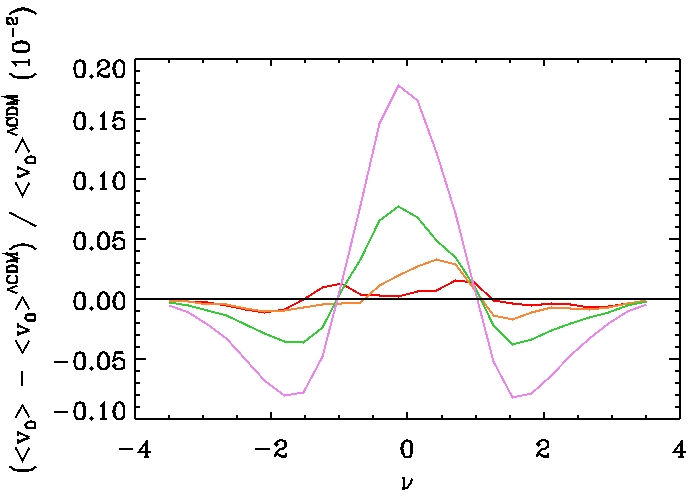}
\includegraphics[scale=0.31]{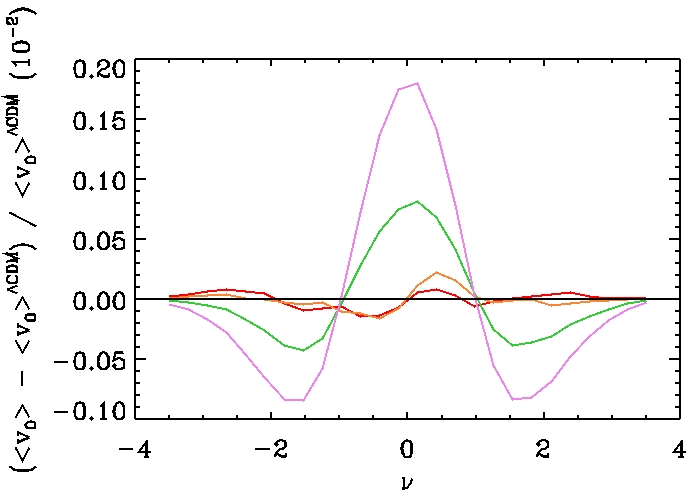} \\
\includegraphics[scale=0.31]{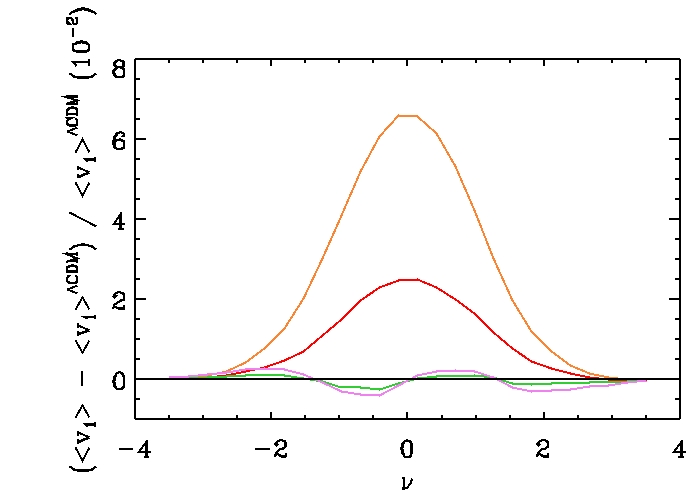}
\includegraphics[scale=0.31]{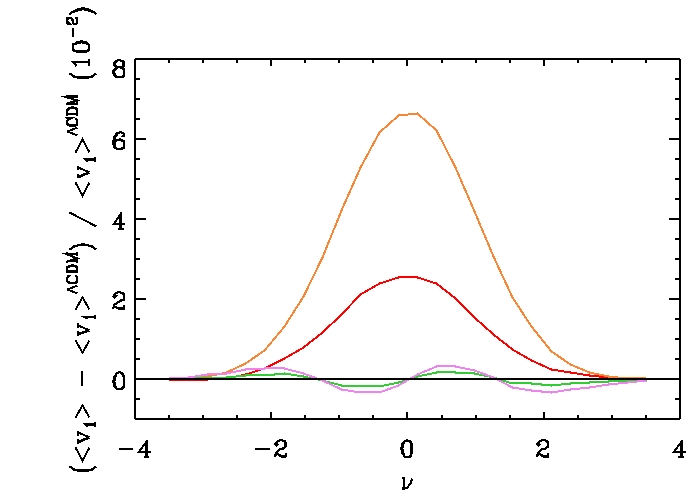} \\
\includegraphics[scale=0.31]{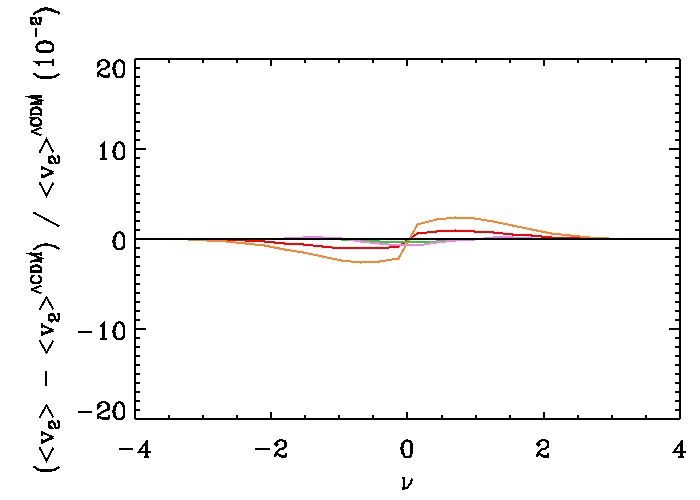}
\includegraphics[scale=0.31]{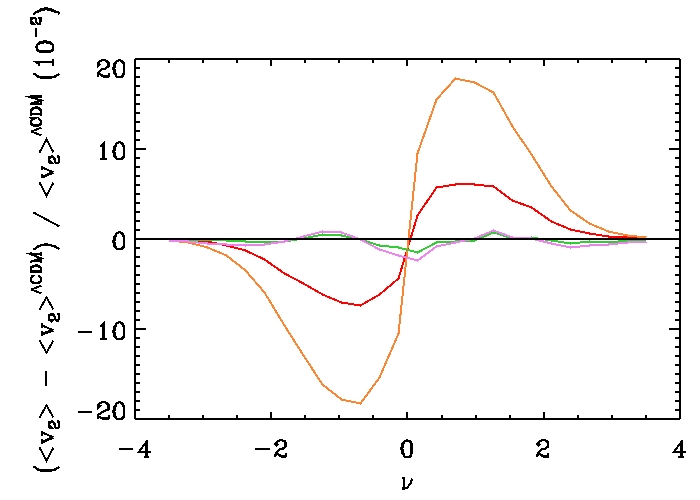} \\
\includegraphics[scale=0.31]{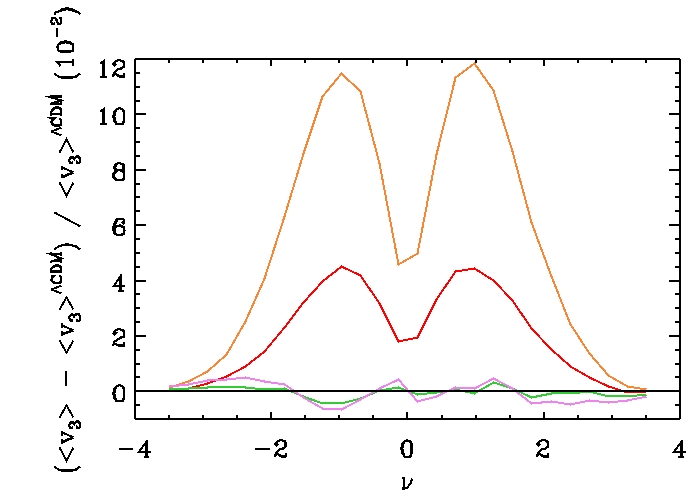}
\includegraphics[scale=0.31]{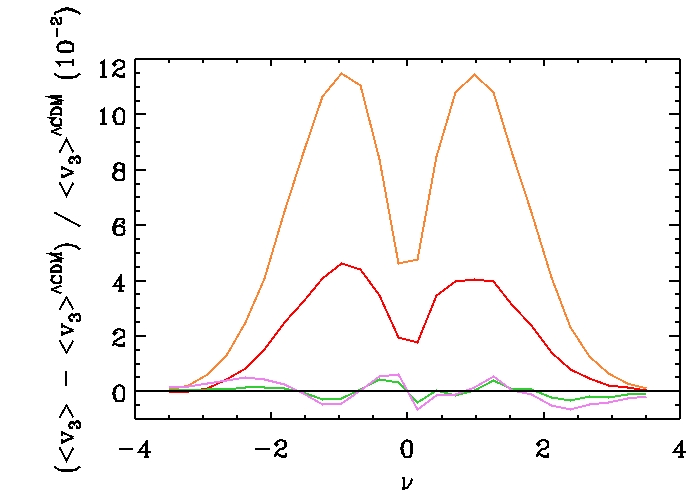}
\caption{
Relative dif\/ference between the mean curves plotted in Fig.~\ref{fig4} of the corresponding MFs 
for the sets of CMB maps (ii) to (v) minus the mean curves obtained for the Gaussian $\Lambda$CDM 
set (i) (see the text for details about these five sets of MC CMB maps). 
The left (right) column refers to analyses with the $\mathtt{UT78}$ mask ($\mathtt{Commander}$-mask), 
to compare the effect left by the use of different masks. 
From top to bottom the four rows of panels refers to the four MFs: Area, Perimeter, Genus, and 
N$_{clusters}$, respectively. 
In each plot the four coloured curves correspond to the dif\/ference between the MFs from simulated 
CMB sets (ii), (iii), (iv), and (v) minus the $\Lambda$CDM set, i.e., the Gaussian MC set (i): 
violet [set (v) - set (i)], green [set (iv) - set (i)], orange [set (iii) - set (i)], and red [set (ii) -set (i)], 
respectively. 
It is worth to notice that the vertical scale is not the same for all plots. 
In the case of the Area-MF, the vertical scale is two orders of magnitude less than the 
other MFs.
} \label{fig5}
\end{figure*}

We show in Fig.~\ref{fig4} our analyses of the four MFs, including a comparison using two masks: 
the $\mathtt{UT78}$ mask (left column) and the $\mathtt{Commander}$-mask (right column). 
From top to bottom we analyse the MFs: Area, Perimeter, Genus, and N$_{clusters}$, respectively. 
In all the panels we show the mean value of the corresponding MF obtained for each of the five 
sets of 1000 MC maps (we illustrate these curves using different colours). 
We also plot the shadow region corresponding to $1\sigma$ values of the Gaussian 
$\Lambda$CDM, i.e., MC set (i). 
At first sight, Area and Genus seems not sensible to the different non-Gaussian contents in the 
sets of MC CMB maps. 
From other side, for the Perimeter and N$_{clusters}$ MFs, and for the Genus case when the 
$\mathtt{Commander}$-mask is used, some curves appear close and indistinguishable. 
For this we find interesting to plot the relative dif\/ference between the mean curves of the MC sets 
(ii) to (v) minus the mean curves obtained from the Gaussian $\Lambda$CDM maps, 
as illustrated in Fig.~\ref{fig5}.

In Fig.~\ref{fig5} the left column refers to analyses with the $\mathtt{UT78}$ mask, while the right 
column refers to analyses with the $\mathtt{Commander}$-mask, just to compare the effect left by the 
use of different masks. 
The four rows of such figure, from top to bottom, refers to the four MFs: Area, Perimeter, Genus, and 
N$_{clusters}$, respectively. 
In each plot the four coloured curves correspond to the relative dif\/ference between the mean curves, 
plotted in Fig.~\ref{fig4}, of the corresponding MFs from simulated CMB sets (ii), (iii), (iv), and (v) 
minus the Gaussian $\Lambda$CDM MC set (i): 
violet [set (v) - set (i)], green [set (iv) - set (i)], orange [set (iii) - set (i)], and red [set (ii) - set (i)], 
respectively. 
It is worth to notice that the vertical scale is not the same for all plots. 
In the case of the Area-MF, the vertical scale is two orders of magnitude less than the 
other MFs. 

The analyses of these results lead to three important conclusions. 
First, a single statistical estimator is not sensible enough to detect any non-Gaussian contribution 
(or a combination of them) present in CMB maps; that is to say: various estimators are more efficient 
to perform this task. 
Second, analyses with different cut-sky masks reveal basically the same information, except for the 
Genus-MF as noticed in the plots of the third row of Fig.~\ref{fig4}. 
Third, and more interestingly, the Gaussian deviations shown in the plots of Fig.~\ref{fig5} makes clear 
that the set of MFs are able to discriminate distinct NGs imprinted in CMB maps: 
the non-Gaussian signatures appearing in the sets of MC CMB maps emerging from {\em Models A} and 
{\em B} are not of the type, neither intensity, expected in maps contaminated with primordial 
{\it local}\/-NG, deserving a closer analyses (to be done in the next subsection).

%-----------------------------  Figura 4  -----------------------
%-----------------------------  Figura 4  -----------------------

%-----------------------------  Figura 5  -----------------------
%-----------------------------  Figura 5  -----------------------

%----------------------------------------------------- B ------------------------------------------------------------o
\subsection{Non-Gaussian features in MC CMB maps from {\em Models A} and {\em B}} 

Our analyses here are aimed to answer the question: are there distinguishable non-Gaussian 
features left in simulated CMB maps generated by {\em Models A} and {\em B}'s angular power spectra? 
For this, we examine the peculiar imprints left in the sets of CMB maps produced from the power 
spectra of {\em Models A}  and {\em B}, features that can be revealed by the MFs in comparison with 
the Gaussian CMB maps generated using the $\Lambda$CDM power spectrum. 
Remember that the power spectra of these {\em Models} were obtained through the CAMB code 
assuming a step-like potential in the inflation model (these {\em Models} differ one to the other in the 
parameter values of the step-like inflationary potential). 
It is also worth to notice that {\em Model A} fits the CMB angular power spectrum from Planck data better 
than the $\Lambda$CDM model does (see Fig.~3 in Ref.~\cite{Benetti:2013cja}, and analysis therein); 
from other side, the angular spectrum from {\em Model B} fits the Planck data worse than the 
$\Lambda$CDM model does. 

Our analyses of the four CMB data sets, that is, the MC sets (ii) to (v), show consistency at a 1$\sigma$ 
level with the MFs obtained from the Gaussian $\Lambda$CDM maps, as seen in Fig.~\ref{fig4}. 
For this, to go further in the current scrutiny we explore the relative dif\/ference curves between the 
mean MF from a given MC set, from set (ii) to set (v), and the corresponding mean 
MF obtained for the set (i), that is the set of Gaussian $\Lambda$CDM CMB maps. 
Our examinations show, except in the Area-MF case, that the CMB maps seeded by the power spectra 
of {\em Models A} (set (ii)) and {\em B} (set (iii)) evidence a larger Gaussian deviation, with fully distinct 
signature, as compared with the results from the MC CMB maps contaminated with primordial 
{\it local}\/-NG, with $f_{\rm \,NL}=38$ (set (iv)) and $f_{\rm \,NL}=100$ (set (v)). 
 
In fact, one verifies in the panels of the second, third, and fourth rows of Fig.~\ref{fig5}, that a 
distinguished departure from Gaussianity comes from the analyses of these CMB data produced from 
{\em Models A} and {\em B}, with the remarkable feature that for the {\em Model B} case the Gaussian 
deviation appears with the same signature than for the {\em Model A} case but it is notoriously larger. 
In other words, inflationary {\em Models A} and {\em B} produce a noticeable (in type and intensity) 
Gaussian deviation as measured by the MFs, where they are best revealed by the Perimeter and 
N$_{clusters}$ MFs, and less efficiently by the Genus, beside being robust under different masks 
applied. 

All these results let us to conclude that the non-Gaussian signals found in the MC maps seeded by the 
CMB power spectra from inflation {\em Models A} and {\em B} are contributions with a definite 
signature originated in the inflationary phase, suggesting an effective probe to the primordial universe.

%----------------------------------------- C -------------------------------------------o
\subsection{Mapping non-Gaussian signatures in Planck maps} 

\begin{figure*}[!h]
\includegraphics[scale=0.33]{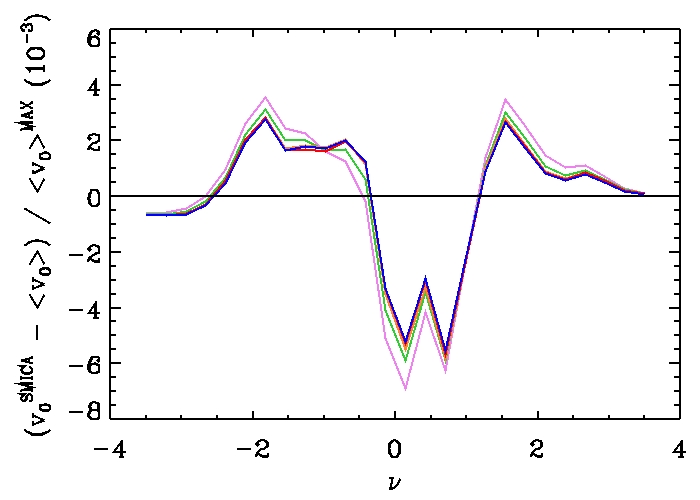}
\includegraphics[scale=0.33]{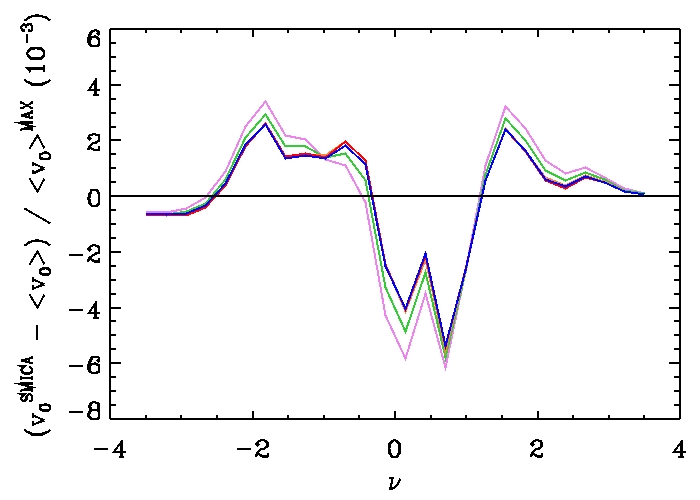} \\
\includegraphics[scale=0.33]{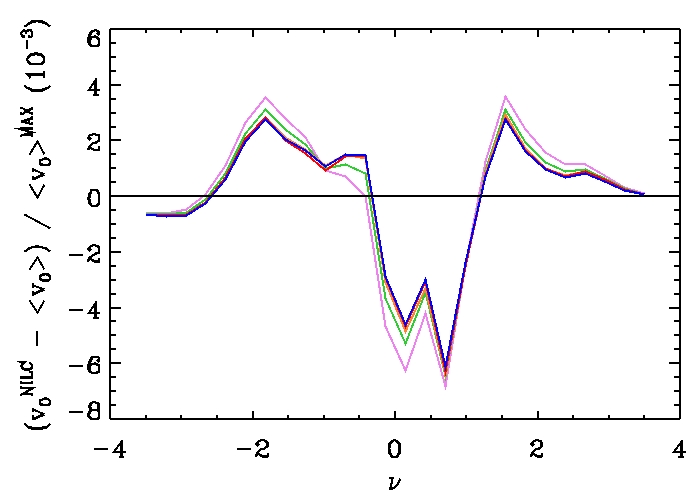}
\includegraphics[scale=0.33]{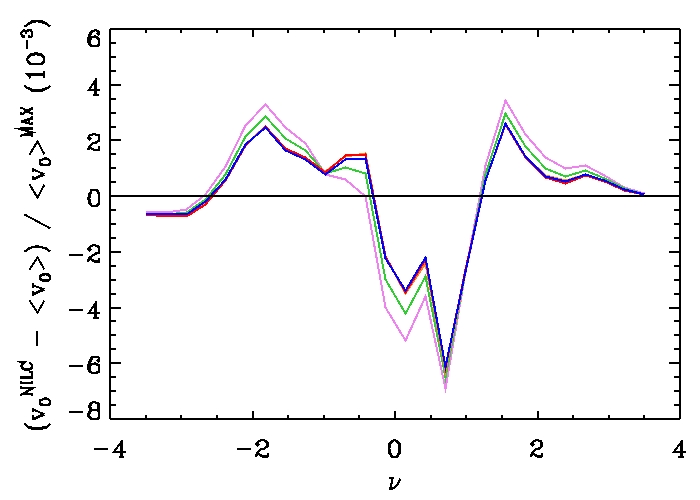} \\
\includegraphics[scale=0.33]{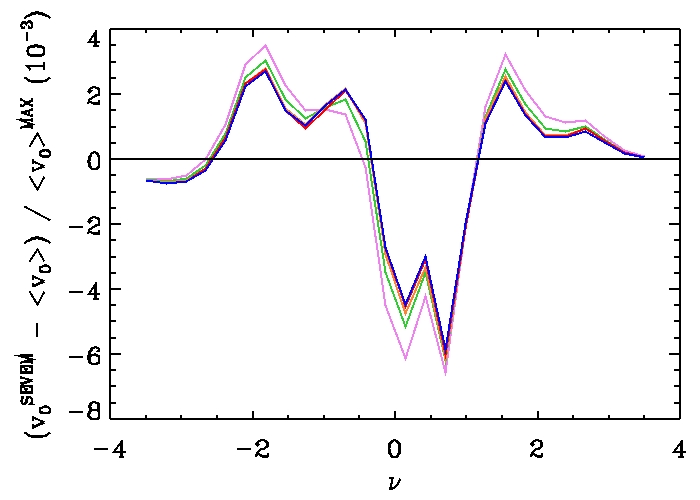}
\includegraphics[scale=0.33]{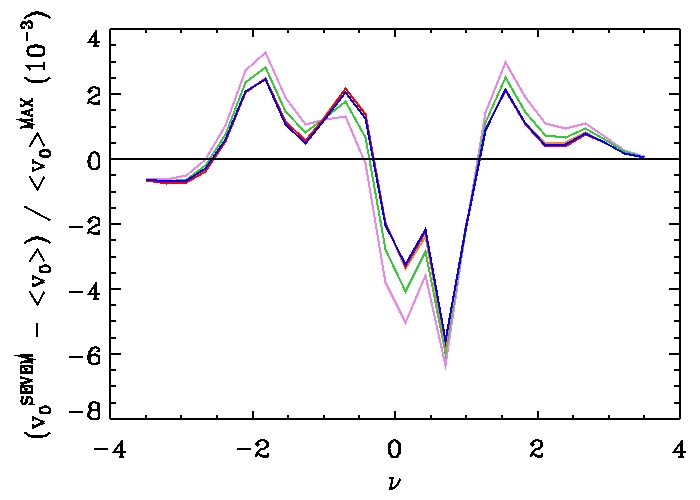} \\
\includegraphics[scale=0.33]{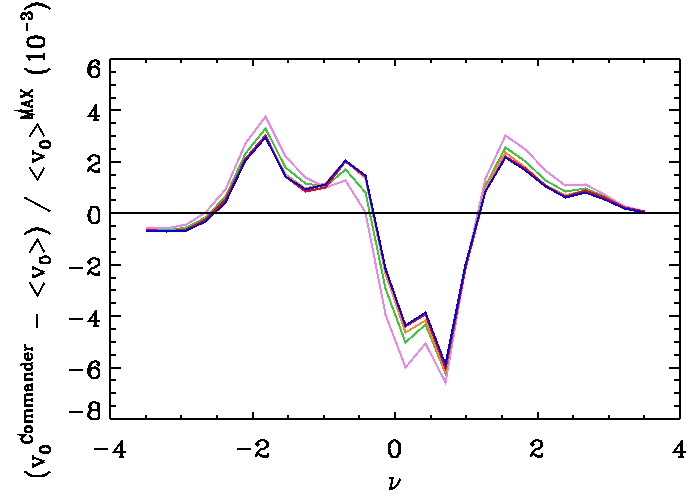}
\includegraphics[scale=0.33]{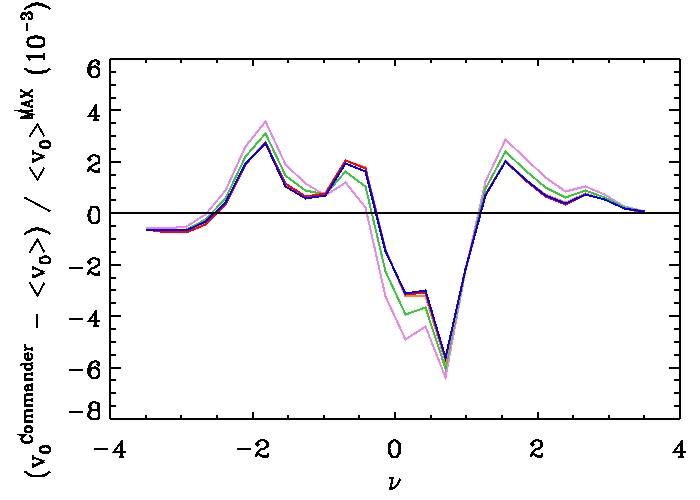}
\caption{\small Relative dif\/ference between the Area-MF vectors ($\mathrm{v}_0$) calculated from each 
Planck map (from top to bottom: $\mathtt{SMICA}$, $\mathtt{NILC}$, $\mathtt{SEVEM}$ and 
$\mathtt{Commander}$ maps, respectively) and the average Area-MF ($\langle \mathrm{v}_0 \rangle$) 
obtained from the five sets of MC CMB maps, namely, $\Lambda$CDM seeded Gaussian (blue) 
and non-Gaussian ($f_{\rm \,NL} = $ 38 and 100; green and violet, respectively) maps, and 
{\em Model A} (red) and {\em Model B} (orange) seeded maps. 
The left and right panels correspond, respectively, to the analyses performed using the masks $\mathtt{UT78}$ 
and $\mathtt{Commander}$-mask.} 
\label{fig6}
\end{figure*}

\begin{figure*}[!h]
\includegraphics[scale=0.33]{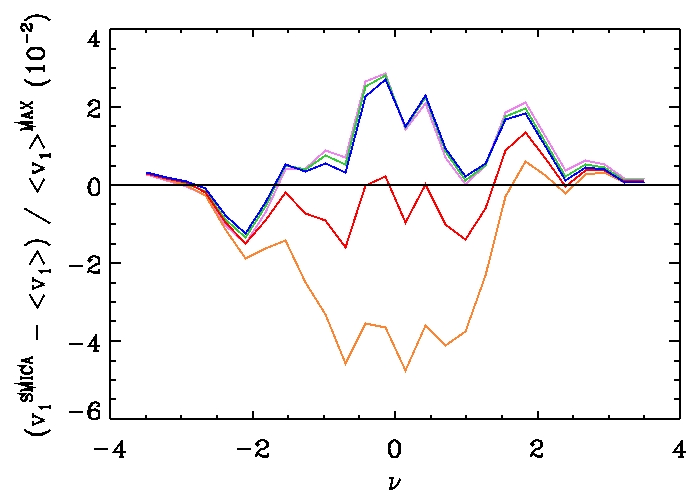}
\includegraphics[scale=0.33]{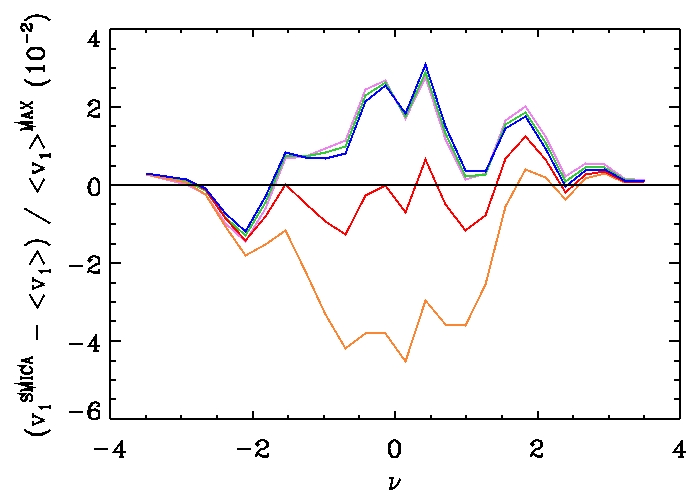} \\
\includegraphics[scale=0.33]{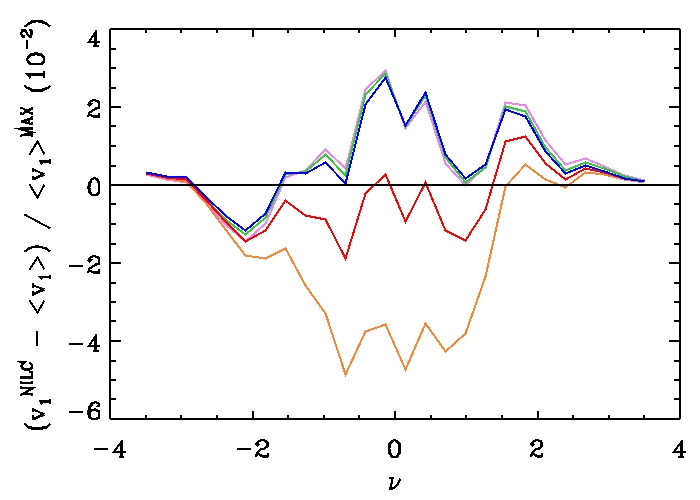}
\includegraphics[scale=0.33]{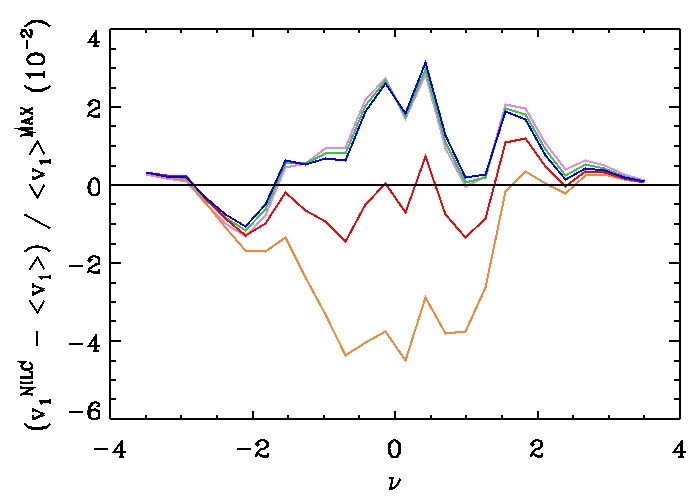} \\
\includegraphics[scale=0.33]{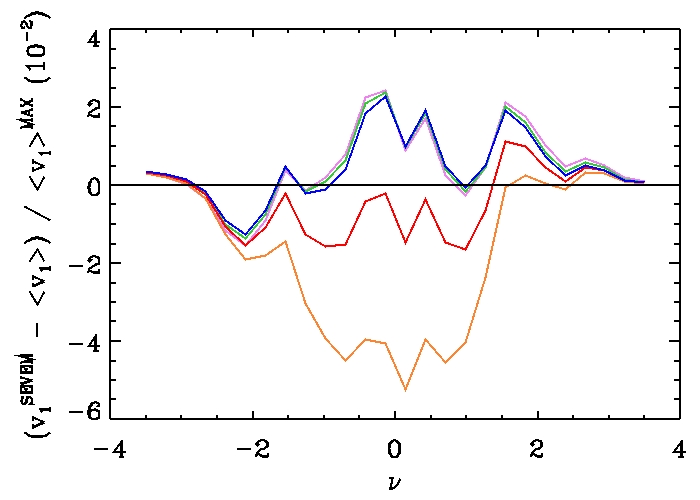}
\includegraphics[scale=0.33]{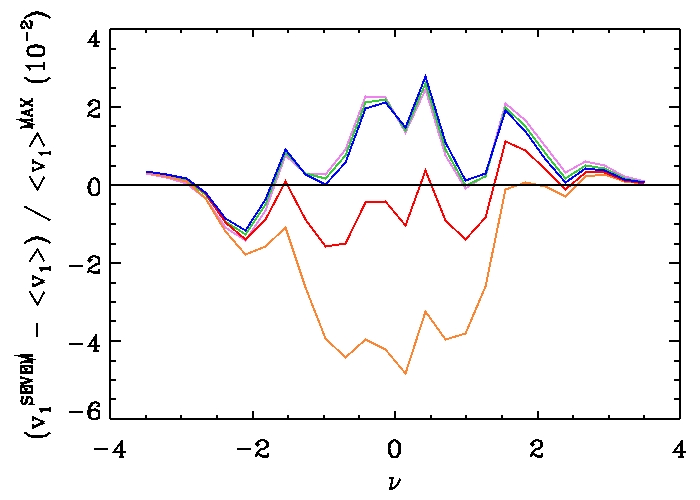} \\
\includegraphics[scale=0.33]{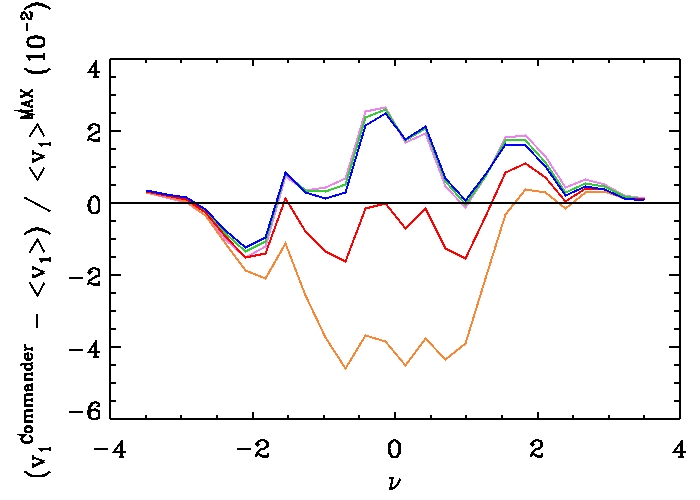}
\includegraphics[scale=0.33]{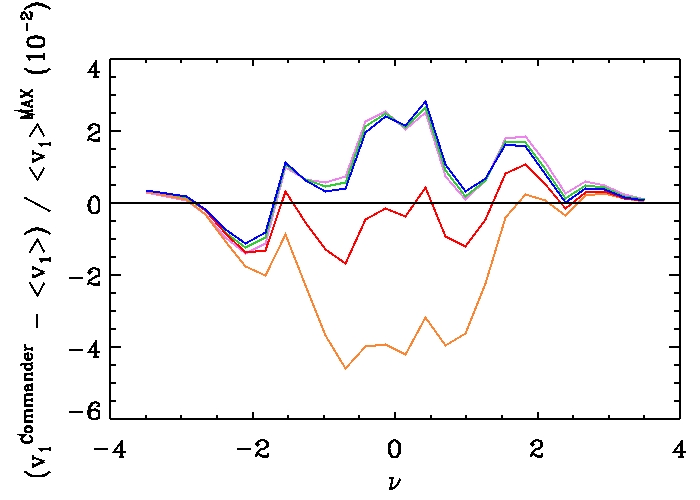}
\caption{\small Same as in Fig. \ref{fig6}, but for the Perimeter-MF vectors.} 
\label{fig7}
\end{figure*}

\begin{figure*}[!h]
\includegraphics[scale=0.33]{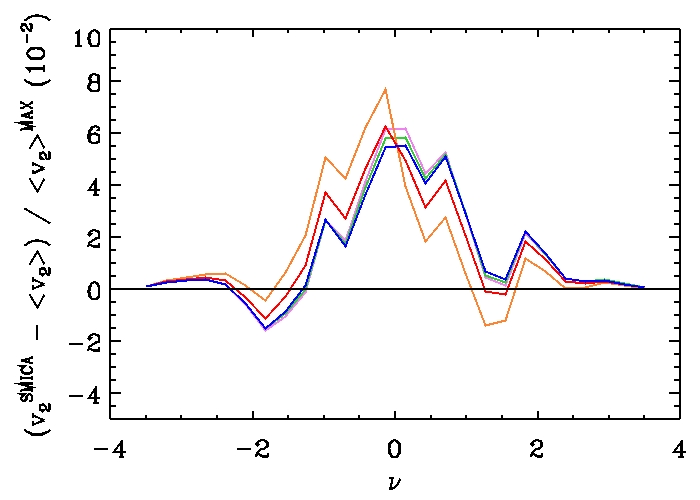}
\includegraphics[scale=0.33]{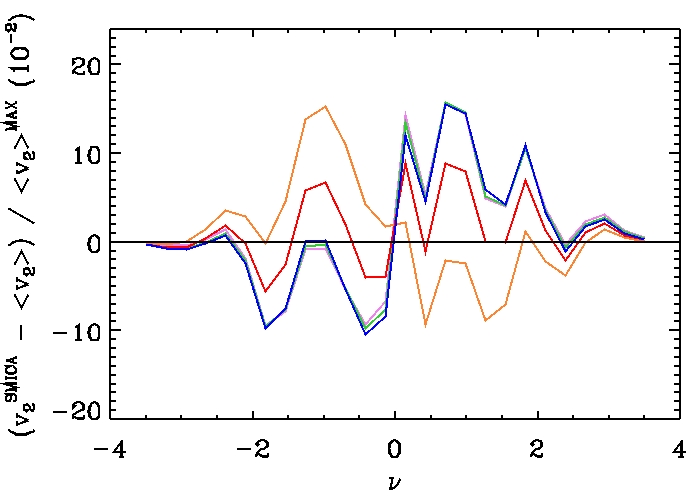} \\
\includegraphics[scale=0.33]{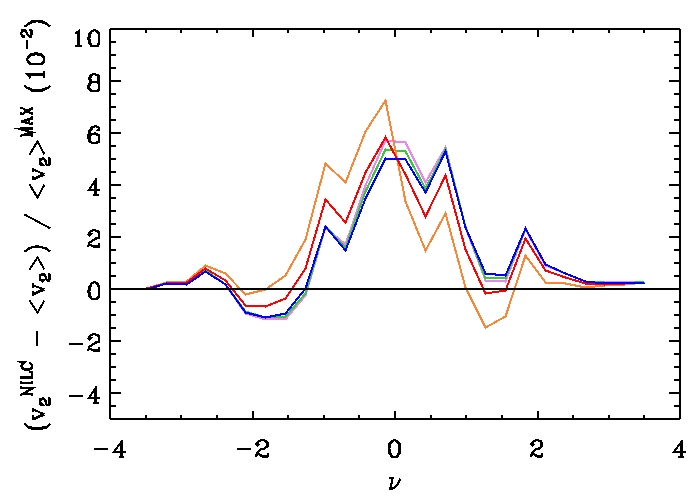}
\includegraphics[scale=0.33]{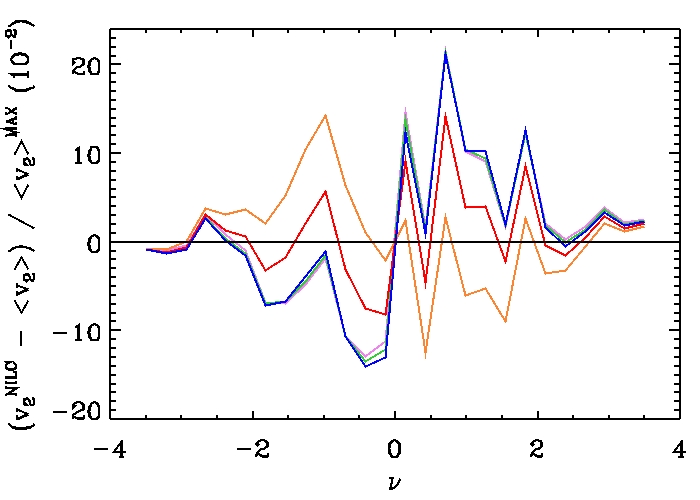} \\
\includegraphics[scale=0.33]{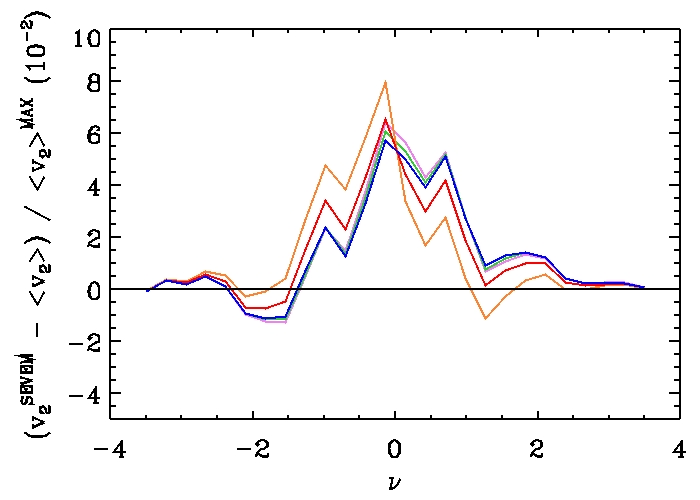}
\includegraphics[scale=0.33]{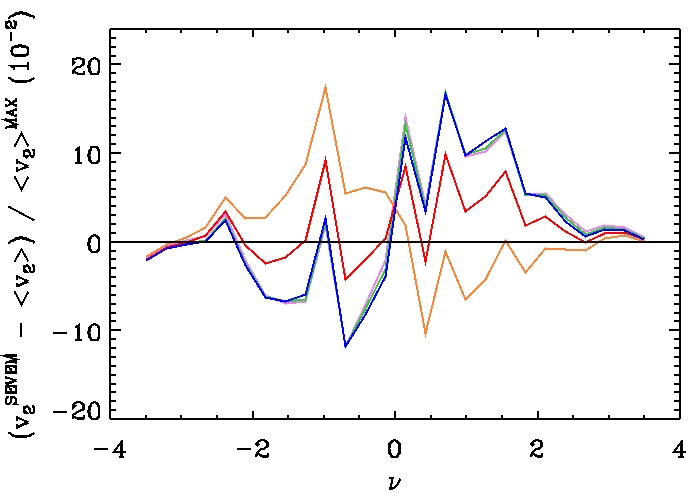} \\
\includegraphics[scale=0.33]{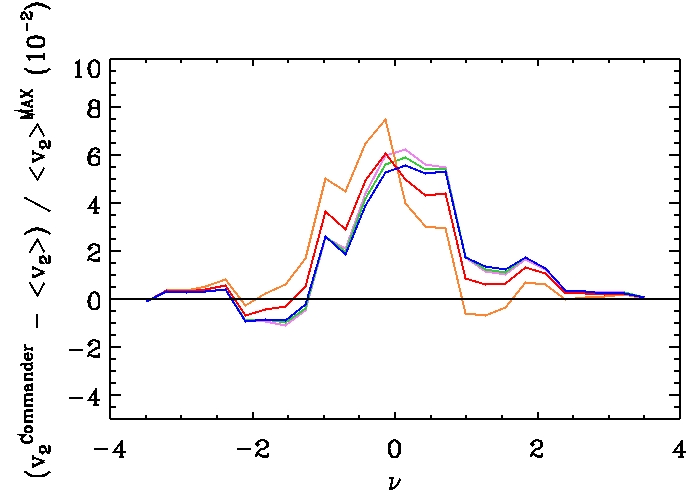}
\includegraphics[scale=0.33]{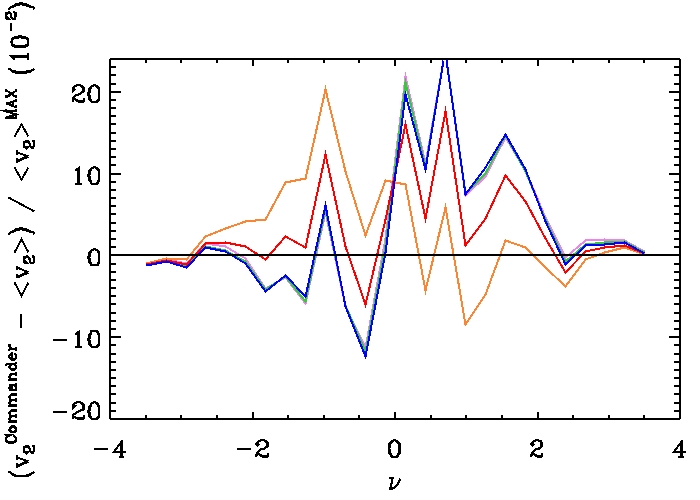}
\caption{\small Same as in Fig. \ref{fig6}, but for the Genus-MF vectors.} 
\label{fig8}
\end{figure*}

\begin{figure*}[!h]
\includegraphics[scale=0.33]{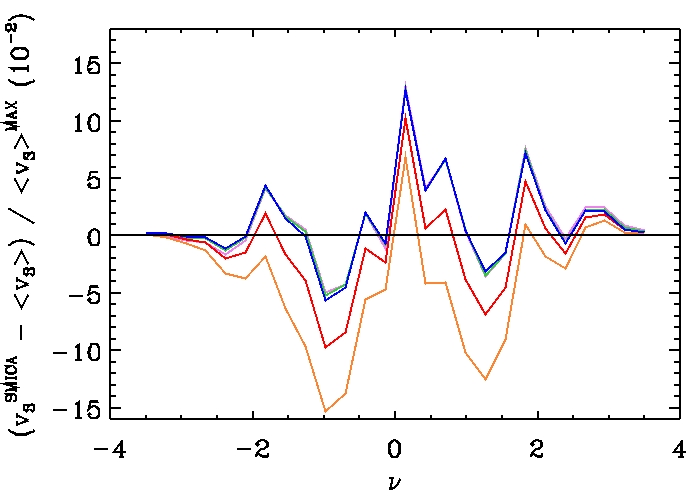}
\includegraphics[scale=0.33]{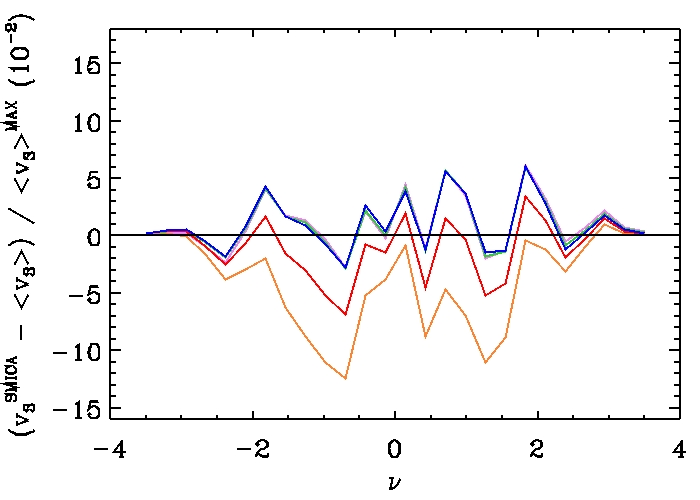} \\
\includegraphics[scale=0.33]{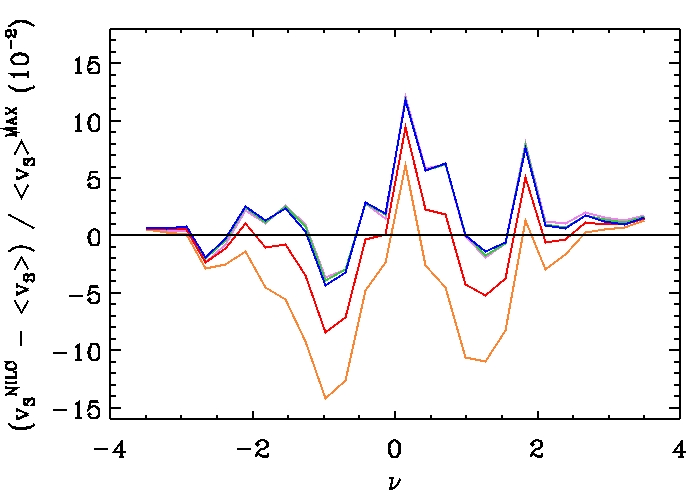}
\includegraphics[scale=0.33]{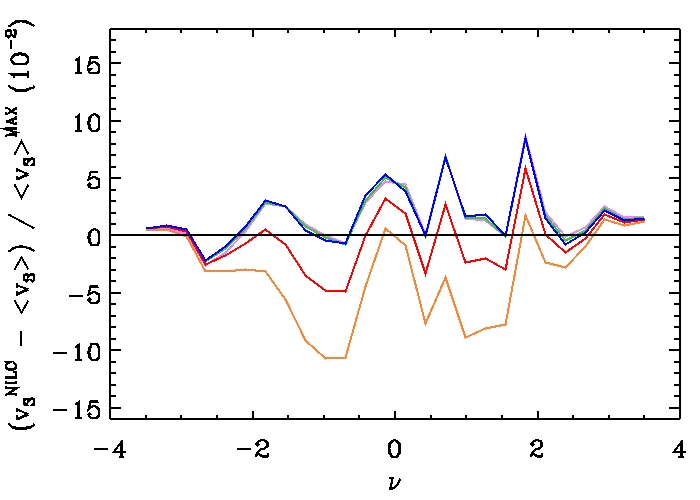} \\
\includegraphics[scale=0.33]{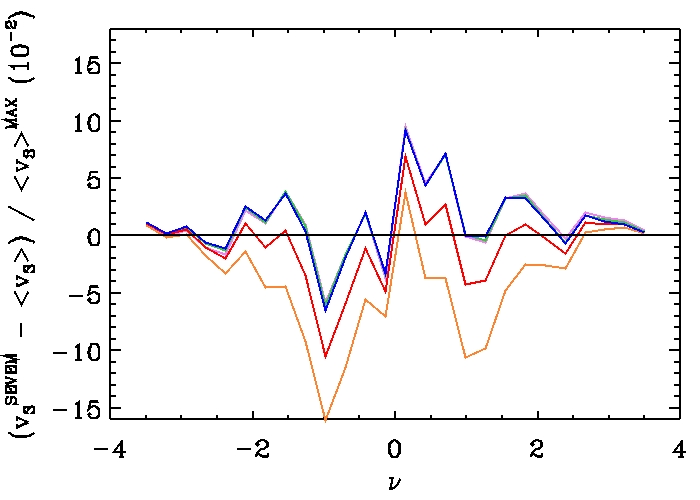}
\includegraphics[scale=0.33]{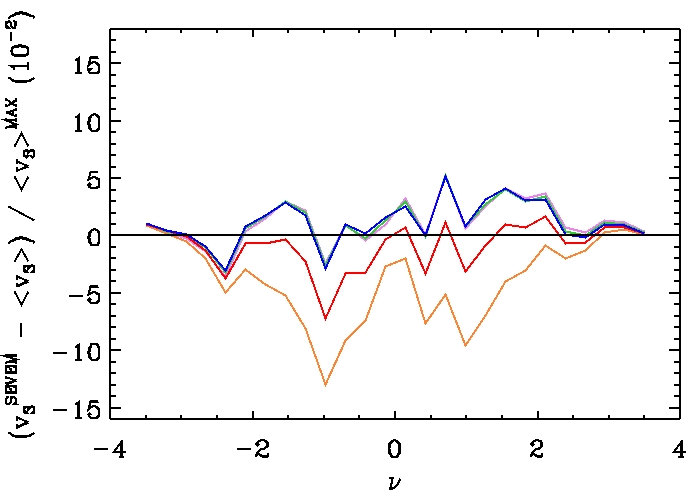} \\
\includegraphics[scale=0.33]{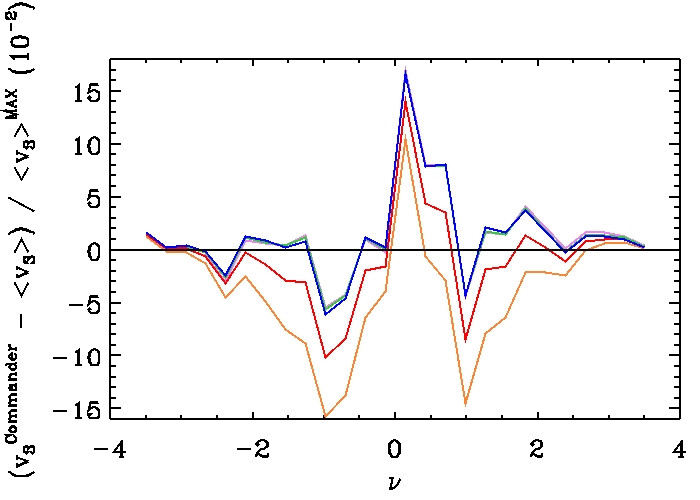}
\includegraphics[scale=0.33]{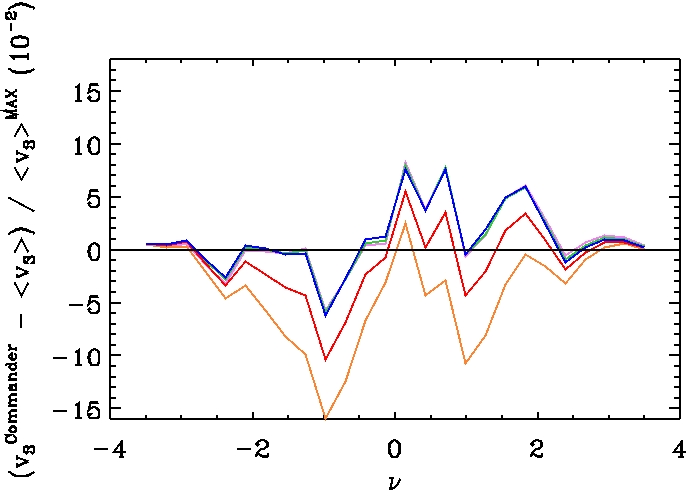}
\caption{\small Same as in Fig. \ref{fig6}, but for the $N_{clusters}$-MF vectors.} 
\label{fig9}
\end{figure*}

The third part of our analyses concerns the calculation of the MFs from the four foreground-cleaned 
Planck CMB maps, considering the above mentioned masks. 
In Figs.~\ref{fig6} to~\ref{fig9} we present the relative dif\/ference between the MFs calculated for 
each one of these Planck maps and the corresponding MF mean obtained for the five sets of MC 
CMB maps. 
In Tables \ref{table1} (for the $\mathtt{UT78}$ mask) and \ref{table2} (for the $\mathtt{Commander}$-mask) 
we show the $\chi^2$ values, for 25 degrees of freedom, corresponding to the best-fit between the MF from 
the Planck map and the mean MF for each of the five MC sets in scrutiny. 
First of all, our results confirm the point raised above regarding the performance between the four MFs. 
As a matter of fact, the Area (Fig.~\ref{fig6}) and Genus (Fig.~\ref{fig8}) functionals appear to be the 
less conclusive, each one presenting quite similar dif\/ferences for all the four Planck maps and for both 
masks. 
Regarding the Perimeter (Fig.~\ref{fig7}) and N$_{clusters}$ (Fig.~\ref{fig9}), the panels show a 
great concordance between Gaussian and non-Gaussian (i.e., $f_{\rm \,NL} \ne 0$) $\Lambda$CDM 
seeded maps, presenting just tiny relative dif\/ferences for the four Planck maps. 
In these figures we also observe that comparing the cases regarding {\em Models A} and {\em B}, they 
exhibit the same signature but very different intensity, being clear that the Perimeter and N$_{clusters}$ 
functionals from the {\em Model A} fits better the four Planck maps. 
Additionally, this last result is corroborated by the $\chi^2$ analyses, being valid for both masks (see 
Tables~\ref{table1} and \ref{table2}). 
In other words, our MF's analyses reveal that the non-Gaussian signatures observed in the four 
foreground-cleaned Planck maps are better represented by those appearing in the set of MC CMB 
maps generated from the {\em Model A} angular power spectrum (i.e., the MC set (ii)). 

\begin{table}[h]
	\centering
	 \caption{$\chi^2$ values calculated from the MFs (Area ($\mathrm{v}_0$), Perimeter 
	 ($\mathrm{v}_1$), 
Genus ($\mathrm{v}_2$), and $N_{clusters}$ ($\mathrm{v}_3$)) curves obtained for each Planck map 
and the average MFs over 1000 realisations of the sets: (i) to (v). 
All the MFs considered here are calculated using the cut-sky corresponding to the $\mathtt{UT78}$ mask. 
The number of degrees of freedom is 25.} 
    \vspace{0.4cm}
    \begin{tabular}{ccccccc}
    \hline \hline
\multirow{2}{*}{\,\,\, MF \,\,\,} & Planck & \multicolumn{5}{c}{Data set$^a$} \\
\cline{3-7}
   & map    & \,\,\,\,(i)\,\, & \,\,\,(ii)\,\, & \,\,\,(iii)\,\, & \,\,\,(iv)\,\, & \,\,\,\,(v)\,\,\,\,\,\, \\ \hline
\multirow{4}{*}{\,\,\, $\mathrm{v}_0$ \,\,\,} 
 & $\mathtt{SMICA}$  & \,\,\,\,34.0 & \,\,\,\,35.9 & \,\,39.8 & \,\,36.7 & 42.1 \\
 & $\mathtt{NILC}$     & \,\,\,\,34.3 & \,\,\,\,36.3 & \,\,40.3 & \,\,37.3 & 43.0 \\
 & $\mathtt{SEVEM}$   & \,\,\,\,33.8 & \,\,\,\,35.6 & \,\,39.6 & \,\,36.0 & 41.0 \\
 & $\mathtt{\,\,Commander\,\,}$ & \,\,\,\,33.0 & \,\,\,\,34.9 & \,\,38.8 & \,\,35.5 & 40.9 \\ \hline
\multirow{4}{*}{$\mathrm{v}_1$} 
 & $\mathtt{SMICA}$ & \,\,\,\,26.2     & \,\,\,\,21.0 & \,\,78.4 & \,\,29.5 & 34.1 \\
 & $\mathtt{NILC}$    & \,\,\,\,27.1     & \,\,\,\,22.9 & \,\,82.3 & \,\,30.5 & 35.3 \\
 & $\mathtt{SEVEM}$   & \,\,\,\,24.8  & \,\,\,\,24.9 & \,\,92.3 & \,\,27.7 & 32.2 \\
 & $\mathtt{\,\,Commander\,\,}$ & \,\,\,\,27.2 & \,\,\,\,23.4 & \,\,83.5 & \,\,30.5 & 35.1 \\ \hline
\multirow{4}{*}{$\mathrm{v}_2$} 
 & $\mathtt{SMICA}$ & \,\,\,\,44.6 & \,\,\,\,43.8 & \,\,47.3 & \,\,48.4 & 51.3 \\
 & $\mathtt{NILC}$    & \,\,\,\,47.7 & \,\,\,\,47.8 & \,\,50.3 & \,\,54.5 & 58.2 \\
 & $\mathtt{SEVEM}$   & \,\,\,\,40.4 & \,\,\,\,39.2 & \,\,43.0 & \,\,43.5 & 45.8 \\
 & $\mathtt{\,\,Commander\,\,}$ & \,\,\,\,44.7 & \,\,\,\,43.5 & \,\,46.7 & \,\,48.0 & 50.4 \\ \hline
\multirow{4}{*}{$\mathrm{v}_3$} 
 & $\mathtt{SMICA}$ & \,\,\,\,22.0 & \,\,\,\,23.5 & \,\,57.9 & \,\,23.7 & 26.3 \\
 & $\mathtt{NILC}$    & \,\,\,\,27.1 & \,\,\,\,26.1 & \,\,55.3 & \,\,29.8 & 32.5 \\
 & $\mathtt{SEVEM}$   & \,\,\,\,18.6 & \,\,\,\,18.4 & \,\,50.3 & \,\,18.7 & 20.1 \\
 & $\mathtt{\,\,Commander\,\,}$ & \,\,\,\,30.0 & \,\,\,\,30.9 & \,\,63.9 & \,\,30.1 & 31.7 \\ \hline \hline
 \multicolumn{7}{p{8cm}}{$^a${\footnotesize The data sets correspond to: MC CMB maps seeded by (i) $\Lambda$CDM, (ii) \textit{Model A}, and (iii) \textit{Model B} angular power spectra, and two other $\Lambda$CDM seeded sets, but including contributions of \textit{local}-NG, namely, (iv) $f_{\rm \,NL} = 38$, and (v) $f_{\rm \,NL} = 100$.}} 
    \end{tabular}
    \label{table1}
\end{table}

\begin{table}[h]
	\centering
	\caption{$\chi^2$ values. 
Same as in Table \ref{table1}, but for MFs calculated using the cut-sky corresponding 
to the $\mathtt{Commander}$-mask. The number of degrees of freedom is 25.} 
	 \vspace{0.4cm}
    \begin{tabular}{ccccccc} 
    \hline \hline
\multirow{2}{*}{\,\,\, MF \,\,\,} & Planck & \multicolumn{5}{c}{Data set} \\ 
\cline{3-7}
   & map    & \,\,\,\,(i)\,\, & \,\,\,(ii)\,\, & \,\,\,(iii)\,\, & \,\,\,(iv)\,\, & \,\,\,\,(v)\,\,\,\,\,\, \\ \hline
\multirow{4}{*}{\,\,\, $\mathrm{v}_0$ \,\,\,} 
 & $\mathtt{SMICA}$ & \,\,\,\,31.9 & \,\,33.5 & 37.5 & 34.0 & 38.9 \\
 & $\mathtt{NILC}$    & \,\,\,\,32.6 & \,\,34.3 & 38.4 & 35.1 & 40.3 \\
 & $\mathtt{SEVEM}$  & \,\,\,\,31.5 & \,\,33.1 & 37.1 & 33.0 & 37.3 \\
 & $\mathtt{\,\,Commander\,\,}$ & \,\,\,\,31.1 & \,\,32.6 & 36.6 & 32.8 & 37.3 \\ \hline
\multirow{4}{*}{\,\,\, $\mathrm{v}_1$ \,\,\,} 
 & $\mathtt{SMICA}$   & \,\,\,\,28.8 & \,\,19.4  & 77.3 & 30.4 & 34.6 \\
 & $\mathtt{NILC}$       & \,\,\,\,29.7 & \,\,21.5 & 82.0 & 31.3 & 35.8 \\
 & $\mathtt{SEVEM}$   & \,\,\,\,27.7 & \,\,22.9 & 89.4 & 28.8 & 32.7 \\
 & $\mathtt{\,\,Commander\,\,}$ & \,\,\,\,29.9 & \,\,22.4 & 83.8 & 31.4 & 35.5 \\ \hline
\multirow{4}{*}{\,\,\, $\mathrm{v}_2$ \,\,\,} 
 & $\mathtt{SMICA}$   & \,\,\,\,20.5 & \,\,10.1 & 18.5 & 20.9 & 22.5 \\
 & $\mathtt{NILC}$       & \,\,\,\,35.2 & \,\,23.8 & 28.1 & 37.4 & 40.2 \\
 & $\mathtt{SEVEM}$   & \,\,\,\,25.6 & \,\,\,\,13.6 & 19.8 & 25.2 & 26.3 \\
 & $\mathtt{\,\,Commander\,\,}$ & \,\,\,\,28.4 & \,\,17.3 & 23.5 & 28.6 & 30.2 \\ \hline
\multirow{4}{*}{\,\,\, $\mathrm{v}_3$ \,\,\,} 
 & $\mathtt{SMICA}$   & \,\,\,\,13.2 & \,\,13.5 & 48.2 & 13.8 & 15.7 \\
 & $\mathtt{NILC}$       & \,\,\,\,27.3 & \,\,23.0 & 49.9 & 28.6 & 31.5 \\
 & $\mathtt{SEVEM}$   & \,\,\,\,13.8 & \,\,12.8 & 45.6 & 14.0 & 15.7 \\
 & $\mathtt{\,\,Commander\,\,}$ & \,\,\,\,18.3 & \,\,19.6 & 55.2 & 18.4 & 20.2 \\ \hline \hline
    \end{tabular}
    \label{table2}
\end{table}

%----------------------------- Figura 6  -----------------------
%----------------------------- Figura 6  -----------------------

%----------------------------- Figura 7  -----------------------
%----------------------------- Figura 7  -----------------------

%----------------------------- Figura 8  -----------------------
%----------------------------- Figura 8  -----------------------

%----------------------------- Figura 9  -----------------------
%----------------------------- Figura 9  -----------------------

\section{Concluding Remarks}\label{sec:conclusion}

The main purpose of this work is the analyses of non-Gaussian signatures originated in the inflation 
models with a step-like feature in the primordial potential. 
Such models produce oscillatory features in the primordial spectrum of scalar perturbations, whose 
effect on the CMB temperature fluctuations can be tested through the analyses of simulated CMB 
maps seeded by the power spectrum generated by this model. 
Here we employed four MFs to search for possible Gaussian deviations in such synthetic CMB maps. 

Using the five sets of MC CMB maps, and examining the effect of using two different cut-sky masks, 
our analyses initiate by testing the sensitivity of the four MFs to {\em local}-NG, with two different 
amplitudes, and that NGs probably produced by the inflationary {\em Models A} and {\em B}. 
A first conclusion is that, between the four quantities, the Genus-MF is the main affected by the cut-sky. 
We also found that the MFs are able to discriminate the different types and amplitudes of NGs. 
This statement is firstly observed from the results presented on Figs.~\ref{fig4} and~\ref{fig5}, which 
display the different features of the {\it local}\/-NG compared to that ones imprinted on CMB by 
{\em Models A} and {\em B}. 
In fact, for the current analyses, we find the following order in sensitivity for the MFs: 
Perimeter $>$ N$_{clusters}$ $>$ Genus $\gg$ Area, from the most to the least sensitive, respectively. 
This result is inferred from the amplitude of the corresponding relative differences observed in 
Fig.~\ref{fig5}, as well as from the $\chi^2$ values shown in Tables \ref{table1} and \ref{table2}. 

The detection of Gaussian deviations, as evidenced by departures of the MFs curves compared with the 
corresponding ones from Gaussian CMB maps, clearly suggest the presence of non-Gaussian signals. 
Exploring this inference more carefully, we analysed the relative difference between the mean MFs 
obtained from the four sets of maps, sets (ii) to (v), and that one corresponding to Gaussian simulations, 
set (i), seeded by the $\Lambda$CDM power spectrum. 
From the obtained results, it is inferred that MC CMB maps generated from {\em Models A} and {\em B} 
power spectra present a clear Gaussian deviation, 
better described by the perimeter-MF analyses as shown in Fig.~\ref{fig7} and quantified by the $\chi^2$ 
values in Tables~\ref{table1} and~\ref{table2}~\footnote{%
Notice that the MFs are not equally sensible to all types of NG (as illustrated, for instance, in the panels 
of Fig.~\ref{fig5}). 
Moreover, the MFs look for all types of NGs, primordial and secondary ones, present in the CMB maps 
in analyses, therefore one does not expect to observe a clear signature from a single contribution, but 
instead a combination of all of them.}, 
whose source is not other than the known `features' 
that characterise these models (see Figs.~\ref{fig2} and \ref{fig3}), therefore originated in the primordial 
inflationary phase. 
Moreover, it was possible to reveal the distinctive signature and amplitude of the Gaussian deviations 
present in the MC CMB maps seeded by the power spectra from {\em Model A}, that one that fits Planck 
data better than $\Lambda$CDM model, and {\em Model B}, 
when compared to the non-Gaussian simulations (i.e., $f_{\rm \,NL} \ne 0$).

Finally, in the last part of this work we analysed the four foreground-cleaned Planck CMB maps in the 
current context. 
The analyses of the relative differences between the MFs of each Planck map and the average MFs 
from the five sets of MC CMB maps confirm a good agreement between the Planck data and 
data coming from the MC seeded by {\em Model A} power spectrum (i.e., the MC set (ii)). 
In addition, we use the goodness-of-fit test to get an overall measure of this concordance 
obtaining the $\chi^2$ values presented in Tables \ref{table1} and \ref{table2}, which confirm this result. 
This lead us to our main conclusion concerning Planck data: 
non-Gaussian signatures present in the four foreground-cleaned Planck maps are better 
represented by the MCs CMB maps seeded by {\em Model A} power spectrum than is done by those 
MCs seeded by the $\Lambda$CDM (Gaussian and {\it local} non-Gaussian simulations) and 
{\em Model B} power spectra. 

Considering that any convincing deviation from Gaussianity results from a robust detection by the 
MFs, which include analyses with different masks. 
In addition, such departures are larger when the power spectrum features appears exaggerated (this 
refers to {\em Model B} `features' with respect to those of {\em Model A}, see Figs.~\ref{fig2} 
and~\ref{fig3}), then the main conclusions of the current analyses 
regarding the MC CMB maps derived from inflationary models with a step-like potential (i.e., the MC 
sets (ii) and (iii)) are: 
\begin{itemize} 
\item 
The non-Gaussian contributions found in these sets of MC CMB maps are of primordial origin. 
\item 
These net Gaussian deviations, as robustly detected by the MFs, are not accounted by primordial 
{\em local}-NG, neither in intensity nor in signature. 
\item
The non-Gaussian signatures in Planck CMB data, as revealed by MFs, are better described by MC 
maps seeded by {\em Model's A} CMB spectrum. 
\end{itemize} 

Additionally, it is opportune to inquire about the type and level of the NG detected. 
According to the results of \textit{Chen et al.}~\cite{Chen06}, inflationary models with a step-like potential 
would produce detectable levels of NG of equilateral type, with expected amplitude 
$| f_{\rm \,NL}^{\rm \,equil} | \simeq 10$, for the step parameters considered in \textit{Model A}.
Moreover, as showed in Table~\ref{table1}, the $\chi^2$ best-fit satisfied by the MCs from \textit{Model A} 
indicates that the NG observed in the foreground-cleaned Planck maps is fully compatible with the NG (of 
equilateral type as predicted by~\cite{Chen06}) found by our perimeter-MF analyses in these MC CMB 
maps (see Fig.~\ref{fig7}). 
These facts, altogether, show the good agreement between the equilateral NG measured by the Planck 
collaboration, namely, $f_{\rm \,NL}^{\rm \,equil} = -16 \pm 70$~\cite{2015/planck-XVII}, and  the NG 
predicted by~\cite{Chen06} for \textit{Model A} and detected by the perimeter-MF in the four Planck CMB 
maps.

%\vspace{-0.4cm}
\section{Acknowledgments}

We acknowledge L.R. Abramo for useful comments. 
CPN and MB acknowledge CAPES and FAPERJ (Brazilian Financial Agencies), respectively, 
for their fellowships; AB acknowledges a CAPES PVE project. 
We are grateful for the use of the $\{ a^{\mbox{\footnotesize G}}_{\ell \, m} \}$ and 
$\{ a^{\mbox{\footnotesize NG}}_{\ell \, m} \}$ simulations \citep{2009/elsner}. 
We also acknowledge the use of CAMB (A. Lewis and A. Challinor: 
http://camb.info/)\footnote{http://lambda.gsfc.nasa.gov/toolbox/tb\_camb\_form.cfm}, 
and for the MF's code~\cite{2012/ducout,2012/gay}. 
Some of the results in this paper have been derived using the HEALPix package~\cite{2005/gorski}.

%

%\newpage

\end{document}